\documentclass{article}

\usepackage{caption}

\usepackage[normalem]{ulem}
\usepackage{arxiv}

\usepackage[utf8]{inputenc} 
\usepackage[T1]{fontenc}    
\usepackage{hyperref}       
\usepackage{url}            
\usepackage{booktabs}       
\usepackage{amsfonts}       
\usepackage{nicefrac}       
\usepackage{microtype}      
\usepackage{lipsum}		
\usepackage{graphicx}
\usepackage{natbib}
\usepackage{doi}
\usepackage{amsmath}
\usepackage{xcolor}
\usepackage{multirow}

\title{Physics-Informed Convolutional Transformer for predicting Volatility Surface}

\date{\today}	

\author{
    Soohan Kim\thanks{The work of S. Kim was supported by URP program by Korea Foundation for the Advancement of Science and Creativity.}\\
	Department of Mathematics\\
	Sungkyunkwan University\\
	\texttt{zeanvszed@g.skku.edu} 
	 \and
	 {\bf Seok-Bae Yun}\thanks{S.-B. Yun has been supported by Samsung Science and Technology Foundation under Project Number SSTF-BA1801-02} \\
	Department of Mathematics\\
	Sungkyunkwan University\\
	\texttt{sbyun01@skku.edu} 
	 \and
 	 {\bf Hyeong-Ohk Bae}\thanks{H. Bae is supported by the Basic Research Program through the National Research     Foundation of Korea (NRF) funded by the Ministry of Education and Technology     (NRF-2021R1A2C1093383)} \\
        Department of Financial Engineering\\
        Ajou University, Suwon\\
	\texttt{hobae@ajou.ac.kr} 
	 \and
	 {\bf Muhyun Lee}\\
	Samsung Securities\\
	11 Seocho-daero 74-gil, Seocho-gu, Seoul\\
	\texttt{moo.lee@samsung.com} 
		 \and
    {\bf Youngjoon Hong}\thanks{The work of Y. Hong was supported by Basic Science Research Program through the National Research Foundation of Korea (NRF) funded by the Ministry of Education (NRF-2021R1A2C1093579) and Korean Government (MSIT) (2022R1A4A3033571).} \\
	Department of Mathematical Sciences\\
	Korea Advanced Institute of Science and Technology\\
	\texttt{hongyj@kaist.ac.kr} 
	 }	

\renewcommand{\shorttitle}{\textit{arXiv} Template}

\hypersetup{
pdftitle={Physics-Informed Convolutional Transformer Architectures for Option Volatility Prediction},
pdfsubject={q-bio.NC, q-bio.QM},
pdfauthor={Soohan Kim},
pdfkeywords={Option Volatility, Physics-Informed Neural Networks, Convolutional LSTM, Attention, Convolutional Transformer},
}

\begin{document}
\maketitle

\begin{abstract}
Predicting volatility is important for asset predicting, option pricing and hedging strategies because it cannot be directly observed in the financial market.
The dynamics of the volatility surface is difficult to estimate.
In this paper, we establish a novel architecture based on physics-informed neural networks and convolutional transformers.
The performance of the new architecture is directly compared to other well-known deep-learning architectures, such as standard physics-informed neural networks, convolutional long-short term memory (ConvLSTM), and self-attention ConvLSTM.
Numerical evidence indicates that the proposed physics-informed convolutional transformer network achieves a superior performance than other methods. 
\end{abstract}

\keywords{Volatility \and Black–Scholes model \and Physics-Informed Neural Networks \and Convolutional LSTM \and Attention Mechanism \and Convolutional Transformer}

\section{Introduction}
Options are financial derivatives that are widely used in hedging portfolios. 
Correctly identifying the price of options has been a topic of ongoing interest for both academics and practitioners. The Black-Scholes equation and its variants have made significant progress in this field, as they explain the relationship between observable variables in the market and option prices \citep{black1972the, black1973the, merton1973theory, garman1983foreign, shinde2012study}. 
The original Black-Scholes equation relies on the unrealistic assumption of a constant volatility.
The volatility of an asset is a degree of variation (or a measure of the uncertainty) about its return in a short period of time. 
The movement of an asset price is unpredictable as most people experienced, which implied its volatility never be constant.
Consequently, numerous attempts have been made to relax and generalize this assumption \citep{hull1987thepricing, heston1993aclosed, derman1994riding, CD02, JT05}.
Differently from asset prices observed in the market, the volatility is not observed, which is a hidden Markov process. Since the volatility cannot be directly obtained from the market observation, it is often derived inversely from option prices observed in the market. The resulting volatility is called the (market) implied volatility.
The implied volatilities calculated in this manner are not constant across all strike prices and maturities available for an option of a single underlying asset at a given time. 
Hence, many studies focus on enhancing the prediction capability of the implied volatility.
One widely used example is the local volatility model, in which volatility is a deterministic function of time and the underlying asset price \citep{derman1994riding, dupire1994pricing}. 
Because the reformulated Black-Scholes equation that reflects the local volatility does not have an explicit form of a solution, numerical methods are often utilized to approximate the volatility function.
{The Black-Scholes model and its associated parameters, which will be utilized in our neural network architecture, are discussed in the appendix.}

{
In the Black-Scholes model, the implied volatility appears as a parameter for matching the theoretical price of an option with its market price. 
Option traders expect different volatility values of the same  underlying asset price based on the strike price and time to maturity of options. 
Since different volatility values can be demonstrated in a 3D plot depending on the strike price and time to maturity, the volatility is referred to as the surface. 
The volatility surface's structure and dynamics are essential for pricing options and hedging strategies.
It is widely known that the number of options quoted in the market varies depending on the type of underlying asset and the market liquidity.
Since an observable and meaningful dataset of options is not well-prepared in general, volatility depending on different maturity and strike prices is not easily derived from the market.
Hence, it is important to shape volatility in the form of a surface with respect to maturity and strike price. 
Moreover, since a smooth form of volatility surface derives an equilibrium price where arbitrage opportunities do not exist, volatility surface plays an important role in option pricing.
}
In previous research on predicting volatilities, they were mainly through the calibration of the local volatility surface or traditional statistical regression models \citep{granger2003forecasting}. Calibration methods create a volatility surface from observed option data that can also satisfy Dupire's equation \citep{avellaneda1997calib, andersen1998the, calib, bondarenko2018calib}, whereas statistical models utilize linear regression techniques such as ARIMA and GARCH \citep{factors, linreg}. 

In this study, we propose {\it five} different neural networks to predict volatility and compare various architectures.
We first consider physics-informed neural networks (PINNs) to solve the inverse problem and to estimate the volatility function of the Black-Scholes equations. 
Since daily volatility surface data can be discretized as a sequence of matrices, the prediction task can be reformulated as a spatiotemporal prediction problem. 
For this purpose, we employ three-dimensional convolution-based architectures to predict the spatiotemporal model as in \citep{malliaris1996using}. 
In scientific computing, a family of PINN architectures has been developed using deep learning to numerically solve forward and inverse partial differential equations \citep{raissi2019physics, PINN001, PINN002, PINN003}.
By exploiting the governing equation and necessary boundary conditions on the target loss function to be minimized, they can learn the nonlinear relationships between inputs and outputs, thereby approximating the solution. 
After PINN developed, this approach has been widely applied in both natural sciences and engineering \cite{PINNa-001, PINNa-002, PINNa-003, PINNa-004}.
As for the second prediction model, we adopt the convolutional long-short term memory (ConvLSTM) network to learn the implied volatility.
The ConvLSTM is a hybrid of convolution operations and the traditional fully connected LSTMs \citep{shi2015convolutional, convlstm001, convlstm002}. 
Since the inputs and hidden feature maps in the network are all regarded as tensors, ConvLSTM is suitable for spatiotemporal prediction tasks. 
In the third architecture, the self-attention ConvLSTM (SA-ConvLSTM) network is further implemented to address self-attention memory modules for each cell in the original network. 
It has successfully improved performance by capturing the long-range dependencies in both the spatial and temporal domains \citep{lin2020selfattention, sa-lstm-01}.
For the fourth approach, convolutional transformer (ConvTF) networks are introduced to utilize the strength of the transformer architecture and attention mechanism in a sequential computation task \citep{liu2021convtransformer, convtf01}. 
Transformer is an encoder-decoder neural network for sequence-to-sequence tasks, which successfully models long-range dependencies in Natural Language Processing.
The great success of transformer motivates scientists to utilize transformer in various scientific areas.
In this paper, we empirically show that the ConvTF architecture provides superior results compared to the vanilla PINN, ConvLSTM, and SA-ConvLSTM algorithms.
To the best of our knowledge, the aforementioned architectures have not been used for prediction of implied volatility with real-world financial data.
As for the last approach, we propose a novel architecture by combining the PINN with the transformer structure.
Applying deep learning and finding statistical relationships via extensive iterations has been criticized for its high complexities, which are often inexplicable. 
Therefore, applying PINN in some areas where the governing physical equations are already well explored is compelling. 
However, mathematical theory and statistical approaches should be considered for a comprehensive understanding and prediction in finance. 
Thus, in this study, we propose a physics-informed convolutional transformer (PI-ConvTF) to predict financial data accurately, namely, volatility. 
We also use the predicted volatility values to infer the corresponding call option prices via the Black-Scholes equation. 
After inferring the call option prices, we further estimate the put option prices corresponding to each strike and maturity pair via the put-call parity relationship, taking into account monthly dividends \citep{hull2003textbook}.
To the best of our knowledge, this is the first study to investigate a neural network that combines ConvTF and PINN for an option-pricing model.
In summary, our main contributions are three-fold.

\begin{itemize}
    \item We formally introduce various neural networks learning an implied volatility surface and predicting option price based on historical S\&P 500 options data.
	\item We show that the five different architectures are able to successfully predict volatility surfaces, and comparative studies are being carried out with an aim of measuring accuracy of the models. 
	\item We empirically show that our novel architecture, PI-ConvTF, outperforms the other architectures. 
\end{itemize}

\section{Related Work}
\label{sec:headings}

\subsection{Physical Approach}
\paragraph{Volatility Surface.}
In the market are traded various options with the same underlying asset but different strike prices and different expirations. 
The implied volatility derived from the Black-Scholes equation and financial data forms discrete points on the volatility surface. 
The local volatility model considers this surface as a function of time and underlying asset prices and proves the existence of a unique volatility function applicable to all options of a single underlying asset \citep{derman1994riding, dupire1994pricing}. 
Many numerical methods utilize the observed data points to interpolate and approximate the volatility surface that also satisfies the modified version of the Black-Scholes equation with time-varying volatility \citep{anwar2018astudy, guo2018local, jin2018reconstruction, cen2011arobust}. 
\paragraph{PINNs as Function Approximators.}
As neural networks present a nonlinear approximation via the composition of hidden layers in various network structures and activation functions, their universal approximation properties can provide an alternative approach for solving differential equations.
PINNs have been used to solve partial differential equations numerically set the coordinates as input nodes and the predicted solutions as output nodes, with the traditional deep neural network hidden layers between them \citep{raissi2019physics}. 
The backpropagation learning algorithm is also used to calculate the gradients of the output nodes with respect to the input nodes, which allows the network to derive the current loss resulting from the equation
\citep{raissi2019physics, lagaris1998artificial}. 
The output values and additional boundary conditions are also imposed on the loss function to improve the approximation accuracy.

\subsection{Neural Network Architectures for Spatiotemporal Prediction}
\paragraph{ConvLSTM.}
ConvLSTM is a recurrent neural network version for spatiotemporal prediction that has convolutional structures in both input-to-state and state-to-state transitions.
The ConvLSTM architecture replaces matrix multiplications in the original fully connected LSTM with convolution operations \citep{shi2015convolutional}. 
ConvLSTM determines the future state of a particular cell in the grid based on the inputs and past states of its local neighbors.
Hence, this architecture and its variants have been applied to numerous spatiotemporal prediction tasks such as precipitation forecasting and video-frame prediction.
\paragraph{SA-ConvLSTM.}
SA-ConvLSTM is a modified version of ConvLSTM, with additional implementation of the self-attention memory module \citep{lin2020selfattention}. 
The self-attention memory module performs self-attention operations on the hidden feature map and the newly designed memory unit in each cell to reflect the global spatial dependency between pixels. The two resulting feature maps are then aggregated. 
The output hidden map directly uses this feature information, whereas the memory unit is updated based on a gating mechanism to preserve the temporal dependency between cells. 
Compared with the other RNN based models, the performances of these SA-ConvLSTM models have noticeable improvements in spatiotemporal prediction tasks such as video-frame and traffic-flow predictions.
\paragraph{ConvTF.}
ConvTF modifies the transformer architecture to perform spatiotemporal predictions by introducing a convolutional self-attention mechanism \citep{liu2021convtransformer}. 
The convolutional self-attention mechanism applies a self-attention operation similar to that of SA-ConvLSTM but to every tensor in the sequence and per generated query tensors. 
Other notable differences compared with the original transformer are the stacked convolution operations used for feature embedding, positional encoding changes, and the use of synthetic feed-forward networks (SFFNs), which are U-net-like series of convolution operations, in the prediction stage. ConvTF achieved a better performance in video-frame interpolation and extrapolation tasks than previous state-of-the-art models.

\section{Methods}
\subsection{Data Preparation} \label{data prep}
We used the daily S\&P 500 Index European call, denoted by the ticker 'SPX', for our dataset from 2004/1/5 to 2021/8/13. The statistics of the entire dataset that we utilized are presented in Table \ref{table:stats} in a format that is similar to Table 1 from \cite{tablestats}. 
More precisely, we set the training and test sets from 2004/1/5 to 2019/12/31 and 2020/1/1 to 2021/8/13, respectively, and the data from the latest 20\% of the training set dates were selected as the validation set.
The same datasets were used for all experiments in this study for fair comparisons across all models.

\begin{table}[htp]
    \centering
    \begin{tabular}{cc|cccc}
        \hline
        \multirow{2}{*}{Moneyness} & \multirow{2}{*}{Type} &
            \multicolumn{4}{c}{Days to Expiration} \\
        & & $< 60$ & $60 - 180$ & $\geq 180$ & Total \\
        \hline
        \multirow{3}{*}{$< 0.94$} & Number of contracts & $154371$ & $54295$ & $34298$ & $242964$ \\
        & Average Price & $(372.64)$ & $(503.72)$ & $(659.00)$ & $(442.36)$ \\
        & Average Bid-Ask Spread & $\{4.51\}$ & $\{3.92\}$ & $\{8.90\}$ & $\{5.00\}$ \\
        \hline
        \multirow{3}{*}{0.94-0.97} & Number of contracts & 175157 & 33151 & 11821 & 220129 \\
        & Average Price & (125.37) & (168.22) & (234.01) & (137.66) \\
        & Average Bid-Ask Spread & \{2.87\} & \{2.04\} & \{5.61\} & \{2.90\} \\
        \hline
        \multirow{3}{*}{0.97-1.00} & Number of contracts & 414233 & 74950 & 26835 & 516018 \\
        & Average Price & (59.87) & (105.71) & (178.32) & (72.68) \\
        & Average Bid-Ask Spread & \{1.42\} & \{1.60\} & \{5.12\} & \{1.64\} \\
        \hline
        \multirow{3}{*}{1.00-1.03} & Number of contracts & 525676 & 104193 & 30941 & 660810 \\
        & Average Price & (18.77) & (62.76) & (143.23) & (31.54) \\
        & Average Bid-Ask Spread & \{0.54\} & \{1.26\} & \{4.64\} & \{0.85\} \\
        \hline
        \multirow{3}{*}{1.03-1.06} & Number of contracts & 371810 & 96451 & 24757 & 493018 \\
        & Average Price & (6.43) & (30.57) & (103.07) & (16.01) \\
        & Average Bid-Ask Spread & \{0.38\} & \{1.00\} & \{4.11\} & \{0.69\} \\
        \hline
        \multirow{3}{*}{$\geq 1.06$} & Number of contracts & 319262 & 169002 & 118009 & 606273 \\
        & Average Price & (3.51) & (12.23) & (37.12) & (12.48) \\
        & Average Bid-Ask Spread & \{0.41\} & \{0.88\} & \{3.22\} & \{1.09\} \\
        \hline
        \multirow{3}{*}{Total} & Number of contracts & 1960509 & 532042 & 246661 & 2739212 \\
        & Average Price & (60.02) & (98.49) & (168.32) & (77.24) \\
        & Average Bid-Ask Spread & \{1.20\} & \{1.46\} & \{4.60\} & \{1.55\} \\
        \hline
    \end{tabular}
    \captionsetup{skip=10pt}
    \caption{Descriptive (summary) statistics for call option data. The total statistic is computed through the summation of the number of contracts and by taking a weighted average with respect to the number of contracts for the average price and average bid-ask spread. It is noteworthy that our raw data did not contain option price observations outside of the arbitrage boundaries.}
    \label{table:stats}
\end{table}

\begin{figure}
    \centering
    \includegraphics[scale=0.4, width=15cm]{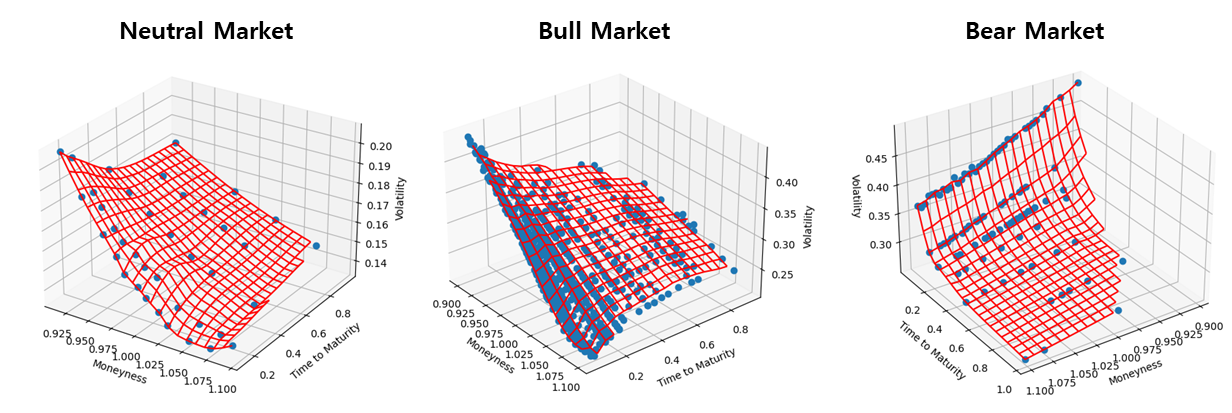}
    \caption{Interpolated volatility surfaces correspond to three market regimes: neutral market, bull market, and bear market. The blue dotted points represent baseline data points from the dataset. When Not-a-Number (NaN) values appear at the boundaries of the grid, these outliers can be replaced by the nearest neighbor values.}
    \label{fig:Vol_Surface_Interpolated}
\end{figure}

In constructing the volatility surface data for each day, we used the corresponding volatility values of quoted options per each moneyness value (we classified the volatility values of each day with their corresponding moneyness value), which is the strike price divided by the underlying asset price, and maturity pair. 
We employ {cubic spline interpolation using the {\it SciPy} interpolation package as in \cite{Boor1978APG}} to generate the surface and sampled discrete points because the raw daily volatility data are not given in a grid-like format over the same moneyness values and maturities. 
We sampled the volatility values of options with 20 moneyness values and maturities ranging from 0.9 to 1.1 and 0 to 1, respectively, using a total of $400$ volatility values given as a $20 \times 20$ matrix per day. 
This choice was based on the characteristics of the most liquid and heavily traded options, as observed from Table \ref{table:stats}.
{After performing cubic spline interpolation, we may encounter Not-a-Number (NaN) values at the boundaries of the grid. These outliers can be replaced with the nearest neighbor values. Figure \ref{fig:Vol_Surface_Interpolated} illustrates examples of interpolated volatility surfaces extracted from different market regimes, including a neutral market, a bull market, and a bear market.}

\subsection{PINN Model}
The PINN model consists of two DNNs:
a network for predicting the call option price $C_{DNN}(S, \tau, m, r)$,  and the volatility function $\sigma_{DNN}(S, \tau, m, r)$.
The inputs of the neural networks are the underlying asset price $S$, time to maturity $\tau$, moneyness value $m$, and risk-free interest rate $r$.

In the traditional PINN setting, the equation to solve typically takes the following form:

\begin{equation}
    \frac{\partial{u}}{\partial{t}} + M[u; \lambda] = 0, \hspace{0.2cm} x \in \Omega, \hspace{0.2cm} t \in [0, T] \label{eq:pinn_fw}
\end{equation}

where $u(t, x)$ is the solution, $M[\cdot; \lambda]$ denotes a nonlinear operator parametrized by $\lambda$, and $\Omega \subset \textbf{R}^{D} (D \in \textbf{N})$. Note that $t$ typically represents time and $x$ is a variable in the spatial domain. $u$ is generally set as a neural network with input data of $t$ and $x$, while Equation \eqref{eq:pinn_fw} gets incorporated in its objective loss function.

In our case, the naive target solution choice would be
\begin{equation}
    u(t, x) = C(\tau, S)
\end{equation}

where $T$ is given as the maximum maturity of our data, i.e. 1 year, and
 
 \begin{equation}
     M[u; \lambda] = M[C; r, \sigma] = - rC + rS\frac{\partial{C}}{\partial{S}} + \frac{1}{2}\sigma^{2}S^{2}\frac{\partial{^2C}}{\partial{S^2}}
 \end{equation}
 
However, this setting treats volatility $\sigma$ as a fixed parameter (constant). Thus, to incorporate the more realistic time-varying volatility in our PINN model construction, we also set $\sigma$ as a target function to solve for. Also, to utilize the strength of neural networks in finding data-driven solutions, we construct both our target solutions $C$ and $\sigma$ as functions of all \textit{observable} variables ($S, \tau, m, r$). In other words, we expand the spatial variable to include $m$ and $r$ as well, i.e. we set $x = (S, m, r) \in \textbf{R}^{3}$. This choice of expansion is a result of the fact that $r$ actually varies in long-term financial data and option volatility is known to vary by its corresponding moneyness value, by a phenomenon referred to as the \textit{volatility smile}.

For simplicity, let 

\begin{equation}
    N(C, \sigma) := \frac{\partial{C}}{\partial{\tau}} + M[C, \sigma]
\end{equation}

Then, the objective loss function that our PINN model aims to minimize can be written as
\begin{equation} \label{e:pinn_loss}
    L_{PINN} = \|\sigma_{DNN} - \sigma_{base}\|_{1} + \|N(C_{DNN}, \sigma_{DNN})\|_{1}
\end{equation}
{where $\sigma_{base}$ is a baseline quantity extracted from the interpolated volatility surface.}

\begin{figure}
    \centering
    \includegraphics[scale=0.35, width=15cm]{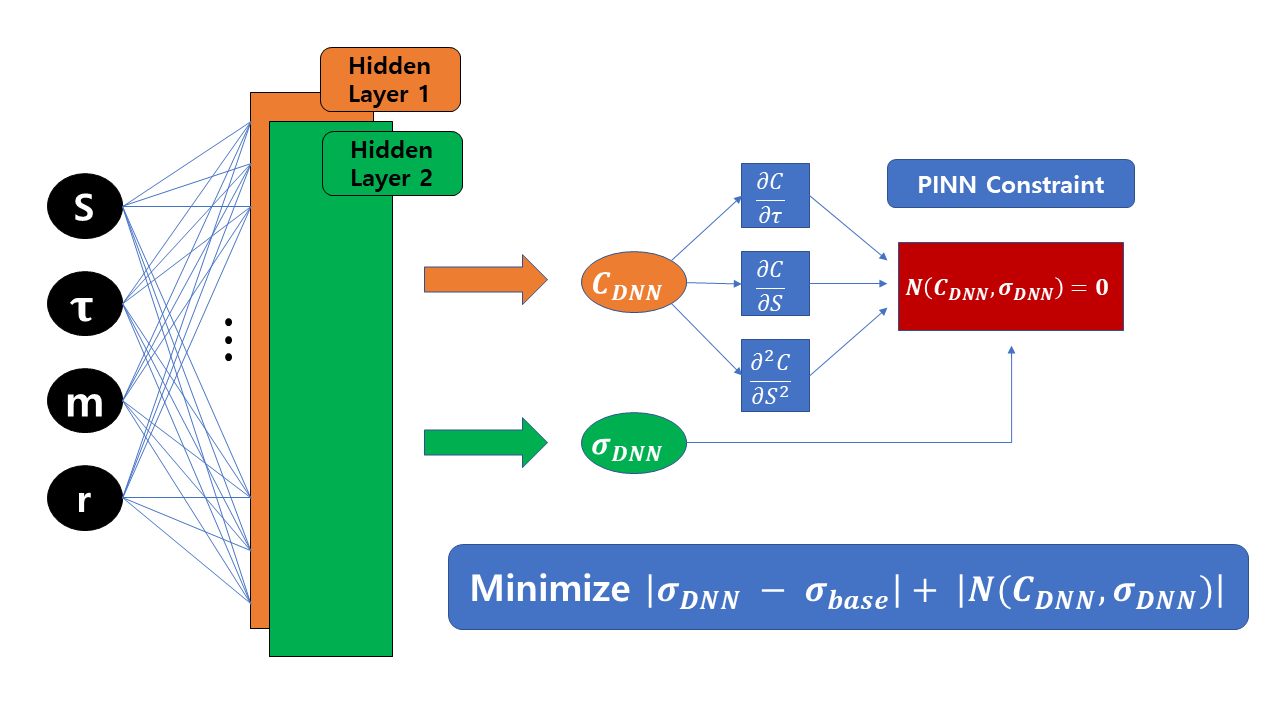}
    \caption{Architecture of the Physics-Informed Neural Network (PINN) model. The diagram showcases the multiple layers and neurons integrated into the network. Starting with the input layer, which processes raw data ($s, \tau, m, r$), the model progresses through a series of hidden layers equipped with activation functions designed for efficient learning. The architecture also emphasizes the incorporation of physical constraints or prior knowledge, which guides the training process and ensures predictions adhere to known physical laws in \eqref{e:pinn_loss}. }
    \label{fig:PINN_Architecture}
\end{figure}

 The structure of the deep neural networks for $C_{DNN}$ and $\sigma_{DNN}$ consists of a single hidden layer with 10,000 nodes and a soft-plus activation function. 
 Figure \ref{fig:PINN_Architecture} shows the detailed architecture of the proposed PINN model.

\subsection{Convolution-based Models} \label{convmodels}

Note that for all convolution-based models, the input tensors are $\{X_{t}\}, (t = 1, 2, \cdots , n)$, where each $X_{t} \in \textbf{R}^{1 \times 20 \times 20}$ is the volatility data of the $t^{th}$ day and $n$ denotes the predefined number of past days to refer to for prediction. 
The ConvLSTM, SA-ConvLSTM, and ConvTF models only utilize this input data, whereas PI-ConvTF also uses other market variable information such as \textit{time to maturity}, \textit{underlying asset price}, \textit{risk-free interest rate}, and \textit{strike price} to additionally impose the Black-Scholes equation. 
In fact, using the diverse set of inputs empirically does not improve the performance of the ConvLSTM, SA-ConvLSTM, and ConvTF models. 
For numerical experiments in PI-ConvTF, we set $X_{t} := [\tau_{t}; \sigma_{t}; S'_{t}; r_{t}; K'_{t}] \in \textbf{R}^{5 \times 20 \times 20}$, where $S'$ and $K'$ denote the normalized underlying asset price and strike price.

\paragraph{ConvLSTM and SA-ConvLSTM Model.}
\begin{figure}[htp]
    \centering
    \includegraphics[scale=0.35, width=\textwidth]{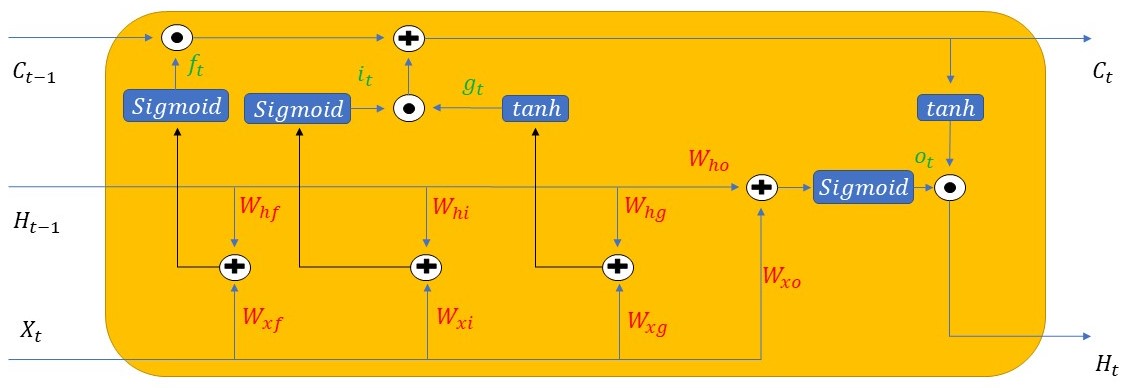}
    \caption{Structure of a Convolutional Long Short-Term Memory (ConvLSTM) cell. Unlike traditional LSTM cells that use fully connected operations, the ConvLSTM employs convolutional operations in both the input-to-state and state-to-state transitions. This diagram illustrates the internal components of the ConvLSTM cell, which include the input gate, forget gate, output gate, and memory cell as in \eqref{e:sa1}--\eqref{e:sa3}. 
    Each gate processes spatial data through convolutional layers, preserving spatial correlations and enabling the cell to handle multidimensional data more effectively.}
    \label{fig:ConvLSTMCell}
\end{figure}

The ConvLSTM model comprises three main functional layers: convolution, flattened, and LSTM. 
We set the number of cells $n$ to the number of previous days, and the number of layers $N$ denotes the number of ConvLSTM layers; the architecture of the ConvLSTM cell is depicted in Figure \ref{fig:ConvLSTMCell}. 
In the first ConvLSTM layer, the inputs of the $t^{th} (t = 1, 2, \cdots , n)$ cell are the current input tensor $X_{t}$ (the volatility surface data of the $t^{th}$ day, of size $20 \times 20$) and the previous cell state $C_{t-1}$ and hidden feature map $H_{t-1}$. Its outputs are the current cell state $C_{t}$ and hidden feature map $H_{t}$.
The compact forms of the equations for ConvLSTM are as follows:

\begin{align}
    & f_{t} = Sigmoid(W_{xf} * X_{t} + W_{hf} * H_{t-1}), 
    & i_{t} & = Sigmoid(W_{xi} * X_{t} + W_{hi} * H_{t-1}), \label{e:sa1}\\ 
    & g_{t} = \tanh(W_{xg} * X_{t} + W_{hg} * H_{t-1}), 
    & C_{t} & = f_{t} \cdot C_{t-1} + i_{t} \cdot g_{t},\label{e:sa2}\\
    & o_{t} = Sigmoid(W_{xo} * X_{t} + W_{ho} * H_{t-1}),
    & H_{t} & = o_{t} \cdot \tanh(C_{t}),\label{e:sa3}
\end{align}

where $*$ denotes the convolution operator and $\cdot$ denotes element-wise multiplication.
We choose $Sigmoid$ as our activation function, and $W_{ConvLSTMCell} = \{ W_{xf}, W_{hf}, W_{xi}, W_{hi}, W_{xg}, W_{hg}, W_{xo}, W_{ho} \}$ denotes a set of learnable convolution kernel weights whose size is configurable for a single cell. The number of channels per weight tensor in $W_{ConvLSTM}$ is selected to match the dimensions of the element-wise addition in Equations \eqref{e:sa1} and \eqref{e:sa2}.
It is also worth noting that in the multi-layered ConvLSTM model, the information of the feature map $H_{layer(i)} = \{ H^{(i)}_{1}, H^{(i)}_{2},  \cdots , H^{(i)}_{n} \}$ is transferred to the input of the next layer, that is, $H_{layer(i)} = X_{layer(i+1)}$. 
The final prediction of the volatility surface on the $(n+1)^{th}$ day can be written as $\sigma_{n+1} = W_{final} * H^{(N)}_{n}$, where $W_{final}$ transforms $H^{(N)}_{n}$ into a tensor of $1 \times 20 \times 20$.

\begin{figure}[htp]
    \centering
    \includegraphics[scale=0.35, width=\textwidth]{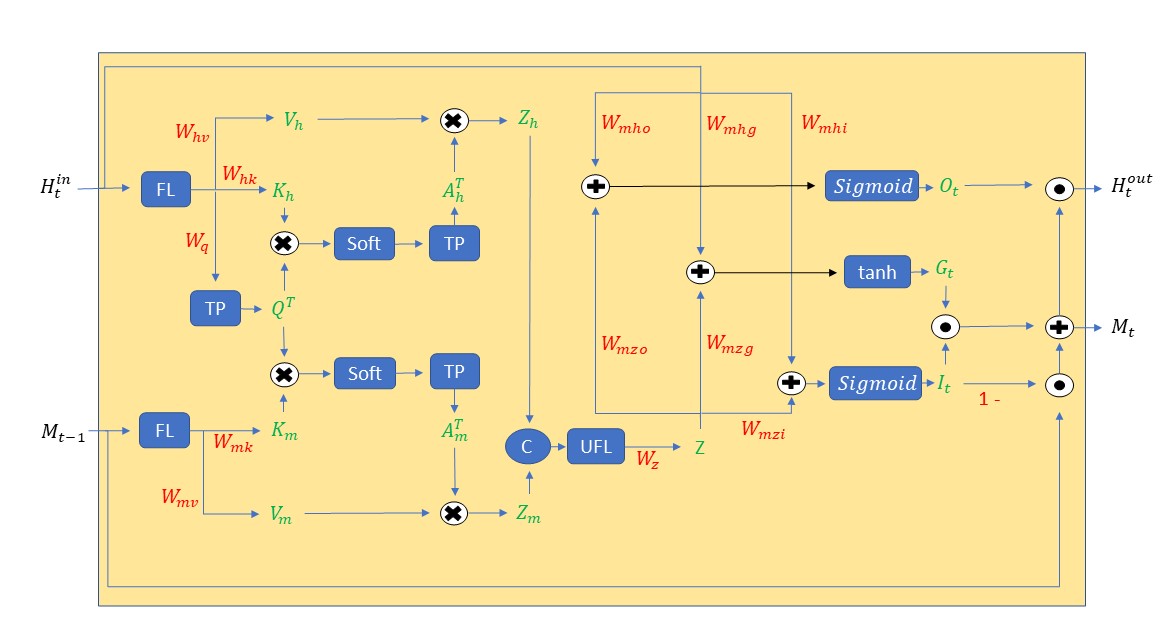}
    \caption{Structure of the Self-Attention Memory Module in the SA-ConvLSTM model. Building upon the foundational architecture of the standard ConvLSTM, this model introduces a unique self-attention memory module designed to capture long-range dependencies across both spatial and temporal dimensions. Within each SA-ConvLSTM cell, this module processes a memory unit tensor $M$, dimensionally consistent with the hidden feature map $H$. It integrates information from the prior memory unit $M_{t-1}$ and the current feature map $H_{t}^{in}$ (derived from equations \eqref{e:sa1}, \eqref{e:sa2}, and \eqref{e:sa3}), producing an updated memory unit $M_{t}$ and refined feature map $H_{t}^{out}$. Subsequent operations remain in line with traditional ConvLSTM procedures.}
    \label{fig:SAModule}
\end{figure}

The structure of the SA-ConvLSTM model is similar to that of ConvLSTM.
Specifically, a novel self-attention memory is proposed to memorize features with long-range dependencies in terms of spatial and temporal domains.
We embed the self-attention memory module into ConvLSTM to construct SA-ConvLSTM; see Figure \ref{fig:SAModule} for a detailed description.
A new tensor, referred to as the memory unit $M$, has the exact dimensions as the hidden feature map $H$.
The self-attention memory module is embedded in each ConvLSTM cell.
The inputs are the memory unit of the previous cell $M_{t-1}$ and the current hidden feature map $H_{t}^{in}$. 
Note that $H_{t}^{in}$ is a tensor obtained at the current ConvLSTM cell following the operations depicted in Equations \eqref{e:sa1}, \eqref{e:sa2}, and \eqref{e:sa3}. 
The outputs of the module are the current memory unit tensor $M_{t}$ and updated version of the current hidden feature map $H_{t}^{out}$. (The self-attention memory module $M_{t-1}$ and the output of the original ConvLSTM cell $H_{t}$ become the inputs, output $M_{t}$, and update $H_{t}$.)
The remaining calculations are the same as those of the original ConvLSTM cell.

The compact forms of these equations can be formulated as follows:

\begin{align}
    Q &= W_{q} * FL(H_{t}^{in}), & K_{h} &= W_{hk} * FL(H_{t}^{in}), \\
    V_{h} &= W_{hv} * FL(H_{t}^{in}), & K_{m} &= W_{mk} * FL(M_{t-1}), \\
    V_{m} &= W_{mv} * FL(M_{t-1}), & A_{h} &= Soft(Q^TK_{h}), \\ 
    A_{m} &= Soft(Q^TK_{m}), & Z_{h} &= V_{h}A_{h}^T, \\
    Z_{m} &= V_{m}A_{m}^T, & Z &= W_{z} * UFL([Z_{h}; Z_{m}]), \\
    O_{t} &= Sigmoid(W_{mho} * H_{t}^{in} + W_{mzo} * Z), & G_{t} &= tanh(W_{mhg} * H_{t}^{in} + W_{mzg} * Z), \\
    I_{t} &= Sigmoid(W_{mhi} * H_{t}^{in} + W_{mzi} * Z), & M_{t} &= (1 - I_{t}) \cdot M_{t-1} + I_{t} \cdot G_{t}, \\
    H_{t}^{out} &= O_{t} \cdot M_{t},
\end{align}

where $Soft$ denotes the softmax function, $FL$ is the flattening operation of the last two dimensions, and $UFL$ is the unflattening operation of the last dimension. 
The set of learnable 1x1 convolution kernel weights is defined as $W_{SA} = \{ W_{q}, W_{hk}, W_{hv}, W_{mk}, W_{mv}, W_{z}, W_{mzo}, W_{mho}, W_{mzg}, W_{mhg}, W_{mzi}, W_{mhi} \}$. 

In our configuration, the channel dimensions of the input tensors are changed thrice. The first change occurs when the query and key tensors are calculated. $W_{q}$ generates a common query tensor for $H_{t}^{in}$ and $M_{t-1}$. $W_{hk}$ and $W_{mk}$ generate the key tensors for $H_{t}^{in}$ and $M_{t-1}$, respectively. Both query and key tensors have different channel dimensions when compared to the original input tensors, whereas the channels of the value tensors obtained from $W_{hv}$ and $W_{mv}$ remain the same. The second change is during the concatenation of $Z_{h}$ and $Z_{m}$. $Z_{h}$ and $Z_{m}$ are obtained as a result of the self-attention operation applied to $H_{t}^{in}$ and $M_{t-1}$, and their channel dimensions are the same as those of the input tensors. Therefore, concatenation doubles channel dimensions. Subsequently, the third change is made, which is through $W_{z}$ that downsizes the channel dimension by half to be the same as that of the input tensors.
Note that in Figure \ref{fig:SAModule}, $TP$ refers to the transpose operation for the last two dimensions and $C$ refers to channel concatenation.

\begin{figure}[htp]
    \centering
    \includegraphics[scale=0.4, width=\textwidth]{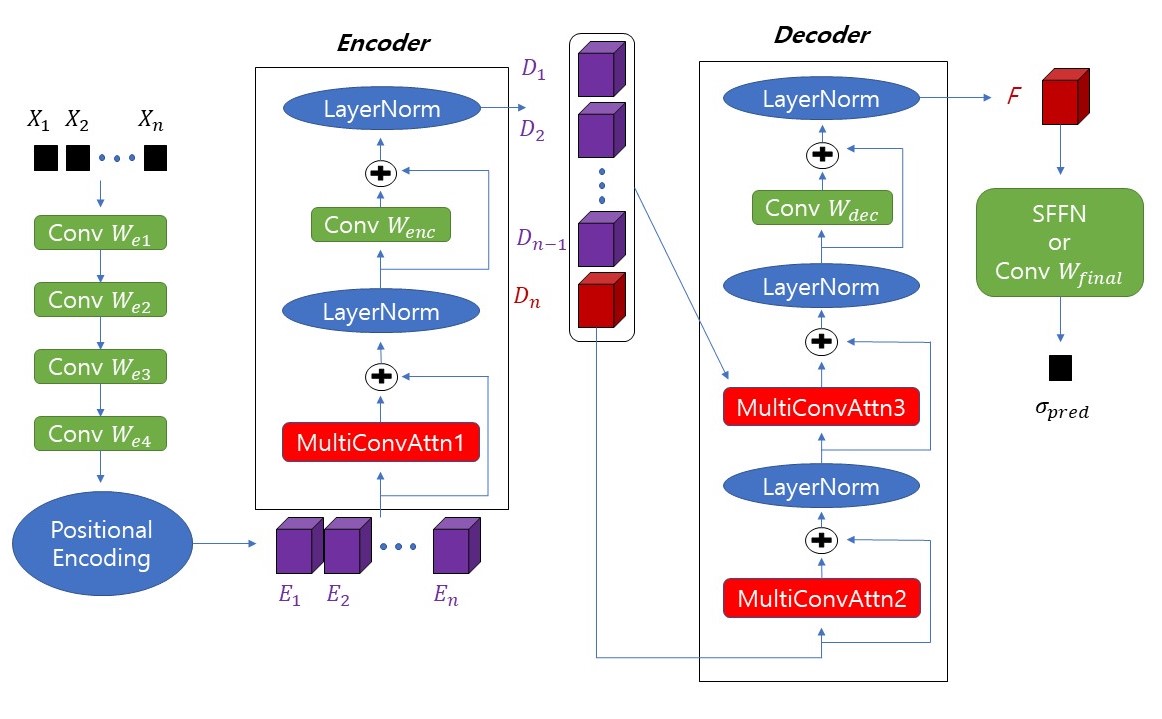}
    \caption{Architecture of ConvTF. An adaptation of the traditional transformer design tailored for vector sequences, ConvTF integrates convolution layers to process sequences of tensors. The model, as delineated in Figure~\ref{fig:ConvTF}, ingests \( n \) input tensors at its embedding layer, along with a single query tensor for the decoder, outputting a tensor subsequently processed by SFFNs or a terminal convolution layer for prediction. Every convolution layer incorporates the \textit{Leaky ReLU} activation function and utilizes \( 3 \times 3 \) convolution kernels, represented as \( W_{e1}, W_{e2}, W_{e3}, W_{e4}, W_{enc}, W_{dec} \), with zero-padding to uphold consistent feature map dimensions across the architecture. The \( \text{MultiConvAttn} \) layer, a multi-head convolutional self-attention mechanism, is pivotal in understanding the long-range dependencies inherent in sequence data. Input tensors are characterized as \( I_j \in \mathbb{R}^{d \times 20 \times 20} \) where the sequence length is \( n \) and the channel count is \( d \).}
    \label{fig:ConvTF}
\end{figure}

\paragraph{ConvTF Model.}
The architecture of ConvTF is a modification of the original transformer architecture for a sequence of vectors. 
Our model configuration of the architecture, described in Figure \ref{fig:ConvTF}, takes $n$ input tensors in the embedding layer and $1$ query tensor input for the decoder and outputs a single tensor that goes through the SFFNs or a final convolution layer to generate the prediction.

In Figure \ref{fig:ConvTF}, all convolution layers are followed by the leaky ReLU activation function, and the weights, $W_{e1}, W_{e2}, W_{e3}, W_{e4}, W_{enc}, W_{dec}$, are all $3 \times 3$ convolution kernels with zero padding to maintain a consistent feature map size throughout the model. 
The weights $W_{e1}, W_{e2}, W_{e3}$, and $W_{e4}$ sequentially increase the channel dimensions of the input tensors. This channel dimension size is retained throughout subsequent operations. Note that the $MultiConvAttn$ layer also outputs tensors with the same channel dimensions, although some changes occur during operations within the layer.

Here, $MultiConvAttn$ denotes the multi-head convolutional self-attention layer that learns the long-range dependence of sequential data.
Let $n$ be the input sequence length, $d$ the number of channels of the input feature maps, and $h$ the number of heads, a divisor of $d$. 
Given an input sequence of tensors $I_j \in \textbf{R}^{d \times 20 \times 20}$ where $1 \leq j \leq n$, and the $3 \times 3$ convolution kernel with zero padding $W^{(j)}_{m}$ where $1 \leq m \leq 3$ and $1 \leq j \leq h$, the $k^{th}$ output of the $MultiConvAttn$ layer is given by

\begin{equation}
    O_{k} = [O^{(1)}_{k}; O^{(2)}_{k}; \cdots; O^{(h)}_{k}], \quad k = 1, 2, \cdots, n.
\end{equation}

Here, $O^{(j)}_{k}$ denotes the output of the $j^{th}$ convolutional attention head, which is written as

\begin{equation}
    O^{(j)}_{k} = \sum_{i=1}^{n} A^{(j)}_{i} \cdot (W^{(j)}_{2} * I_{i}),
\end{equation}

where the $i^{th}$ attention map $A^{(j)}_{i}$ is computed as follows:

\begin{equation}
    A^{(j)}_{i} = Softmax([H^{(j)}_{1}; H^{(j)}_{2}; \cdots; H^{(j)}_{n}])_{(i, \cdot, \cdot)}
\end{equation}

\begin{equation}
    H^{(j)}_{i} = W^{(j)}_{3} * [W^{(j)}_{1} * I_{k}; W^{(j)}_{2} * I_{i}]
\end{equation}

Within these operations, the query (Q), key (K), and value (V) tensors are as follows:

\begin{equation}
    Q^{(j)}_{k} = W^{(j)}_{1} * I_{k}, \quad K^{(j)}_{i} = V^{(j)}_{i} = W^{(j)}_{2} * I_{i}.
\end{equation}

Weights $W^{(j)}_{1}$ and $W^{(j)}_{2}$ are both composed of $d/h$ channels, whereas $W^{(j)}_{3}$ has a single channel for $1 \leq j \leq h$. 
Note that $MultiConvAttn3$ extracts the query tensor from the output of $MultiConvAttn2$, which is a single tensor, and the key and value tensors are obtained from $D_{1}, D_{2}, \cdots, D_{n}$. Thus, the output of the decoder is a single tensor.
We make use of positional encoding as in \citep{liu2021convtransformer}, which introduces ConvTF for the first place. 
As for multi-layered encoders and decoders, the architecture is similar to that of multi-layered ConvLSTMs. 
In the case of encoders, the input of the next layer is the previous layer's output. 
For decoders, the input to $MultiConvAttn2$ is the output of the previous layer, and $MultiConvAttn3$ receives the same exterior inputs $D_{1}, D_{2}, \cdots, D_{n}$ commonly across all layers. 
We implement an SFFN structure consisting of $30$ convolution layers, which repeats gradually widening the number of channels to $128$ and decreasing them to $1$ where $W_{final}$ is chosen to be a $1 \times 1$ convolution kernel with a single channel.

\subsection{PI-ConvTF Model}
\begin{figure}[htp]
    \centering
    \includegraphics[scale=0.4, width=\textwidth]{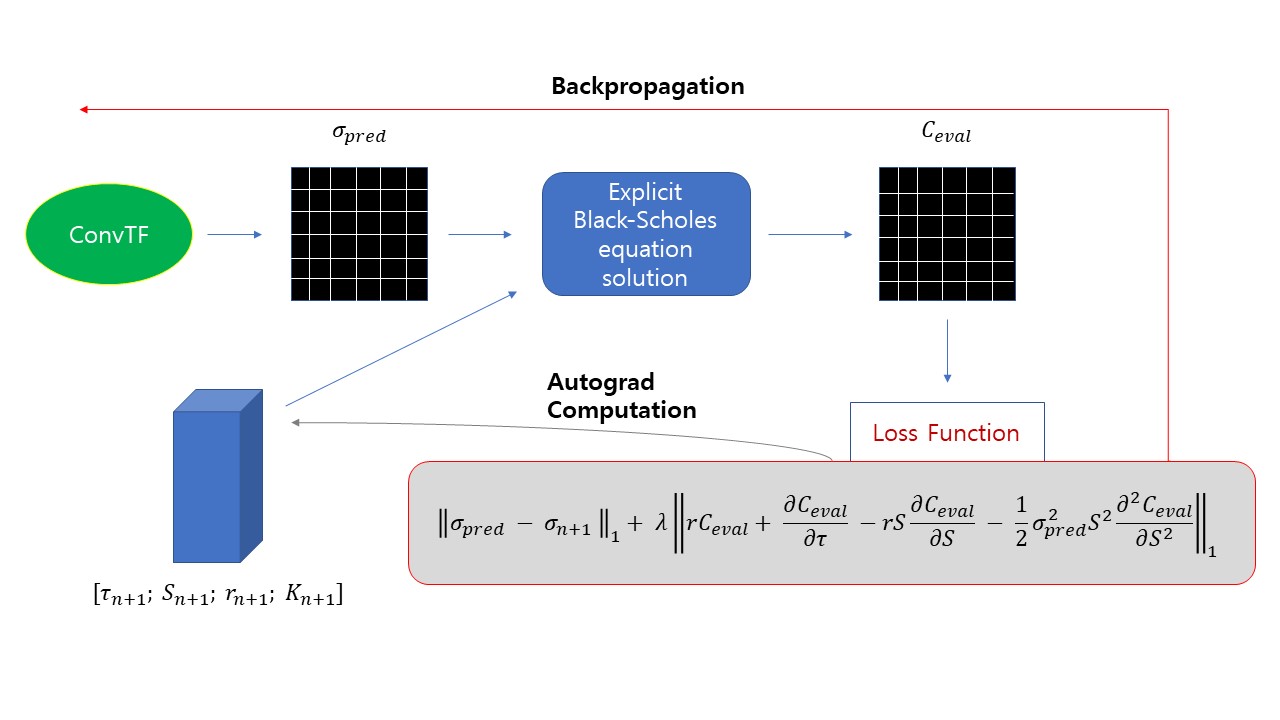}
    \caption{Architecture of PI-ConvTF. The architecture integrates predictions with actual market data on the $(n+1)^{th}$ day: $[\tau_{n+1}; S_{n+1}; r_{n+1}; K_{n+1}] \in \mathbb{R}^{4 \times 20 \times 20}$, where the components represent matrices of \textit{time to maturity}, \textit{underlying asset price}, \textit{risk-free interest rate}, and \textit{strike price}. This is used to compute $C_{eval}$, the call option price for each maturity-strike pair based on the volatility values predicted by ConvTF. The objective loss function for PI-ConvTF then takes into account the actual target volatility, denoted $\sigma_{n+1}$, for the $(n+1)^{th}$ day, and utilizes $\lambda$, an external hyperparameter, to weight the physics-informed loss.}
    \label{fig:PI-ConvTF}
\end{figure}

We propose a novel PI-ConvTF architecture, as shown in Figure \ref{fig:PI-ConvTF}, based on the PINN and ConvTF models. 
In Figure \ref{fig:PI-ConvTF}, $\sigma_{pred}$ denotes the prediction of the volatility surface on the $(n+1)^{th}$ day generated by ConvTF. 
The prediction is then combined with the true data on the $(n+1)^{th}$ day observed from the market $[\tau_{n+1}; S_{n+1}; r_{n+1}; K_{n+1}] \in \textbf{R}^{4 \times 20 \times 20}$, where each component denotes the matrix of the values of \textit{time to maturity}, \textit{underlying asset price}, \textit{risk-free interest rate}, and \textit{strike price}. The matrix $C_{eval}$ is computed as follows:

\begin{equation} 
\begin{split}
    C_{eval_{(i, j)}} = S_{n+1_{(i, j)}}\Phi(d_{1_{(i, j)}}) - K_{n+1_{(i, j)}}e^{-r_{n+1_{(i, j)}}\tau_{n+1_{(i, j)}}}\Phi(d_{2_{(i, j)}}),
\end{split} \label{e:piconvtf1}
\end{equation}
for $1 \leq i,j \leq 20$, where $\Phi$ is the cumulative distribution function of the normal distribution and

\begin{equation} 
\begin{split}
    & d_{1_{(i, j)}} = \frac{ln(S_{n+1_{(i, j)}}/K_{n+1_{(i, j)}}) + (r_{n+1_{(i, j)}} + \frac{1}{2}\sigma_{pred_{(i, j)}}^2)\tau_{n+1_{(i, j)}}}{\sigma_{pred_{(i, j)}}\sqrt{\tau_{n+1_{(i, j)}}}},\\ 
    & d_{2_{(i, j)}} = d_{1_{(i, j)}} - \sigma_{pred_{(i, j)}}\sqrt{\tau_{n+1_{(i, j)}}}.
\end{split} \label{e:piconvtf2}
\end{equation}

Note that \eqref{e:piconvtf1} and \eqref{e:piconvtf2} denote the solutions of the traditional Black-Scholes equation with the hypothesis of constant volatility \citep{black1973the}. 
In our study, we compute the call option price for each maturity-strike pair using the volatility values predicted by ConvTF. This approach is rooted in the real-world practice where implied volatility for each maturity-strike pair is often determined based on equations \eqref{e:piconvtf1} and \eqref{e:piconvtf2} and the observed market price of the corresponding call option. Specifically, practitioners utilize the explicit solution of the Black-Scholes equation to reverse-calculate the implied volatility given an option price, maturity, and strike price.

After $C_{eval}$ is computed, the objective loss function of PI-ConvTF is calculated, as shown in Figure \ref{fig:PI-ConvTF}. 
The real target volatility data of the $(n+1)^{th}$ day are denoted by $\sigma_{n+1}$ and $\lambda$ is an external hyperparameter used for the weights of the physics-informed loss. Gradients for the PINN loss defined as 
\begin{equation*}
    \frac{\partial{C_{eval}}}{\partial\tau}, \quad \frac{\partial{C_{eval}}}{\partial{S}}, \quad \frac{\partial^2{C_{eval}}}{\partial{S^2}}
\end{equation*}
are evaluated using the \textit{autograd} mechanism provided by \textit{PyTorch}.
Because of the direct and indirect (through $C_{eval}$) implementation of $\sigma_{pred}$ in the physics-informed loss, backpropagation successfully updates the weights of ConvTF that are used to compute $\sigma_{pred}$ to minimize the PINN loss.

\section{Experiments and Analysis} 
As described in subsection \ref{data prep}, the training and validation datasets were used to train and validate five different models: PINN, ConvLSTM, SA-ConvLSTM, ConvTF, and PI-ConvTF. Using the mean average percentage error (MAPE) metric, all models were then evaluated on the test dataset. We first performed architecture tuning on each model to determine their peak performance. Subsequently, we provide an overall comparison of the models under their respective best settings to understand the contributions of both physical and statistical approaches to volatility prediction. We also infer the call option price values through Equations \eqref{e:piconvtf1} and \eqref{e:piconvtf2} using the predicted volatility surfaces and check their accuracy to determine the practicality of our methods. {To compare the prediction accuracy of the neural network models with standard statistical methods, we will use Vector Autoregression (VAR) and Autoregressive Integrated Moving Average (ARIMA) to predict multivariate time-series data. This will allow us to evaluate the relative performance of the neural network models in this context.
For VAR, we created input vectors for each day using the point-wise values of the volatility surface. These input vectors were used to predict the volatility values for future dates.
As for ARIMA, we used past data from the same point on the volatility surface as input to predict the point-wise values of the volatility surface.
}

We retrained our models specifically to further develop our findings in the original splitting of the training, validation, and test dates. They were never given the data of historically volatile regimes. Then, they were tested under these regimes. In this way, we experimented with how different models perform under disadvantageous data settings.

\subsection{Implementation}
Because the PINN model attempts to approximate the universal volatility function from daily data points, it was trained more extensively than the other models. The batch size was chosen to be significantly larger because the inputs sampled were not tensors representing the daily volatility surface, as in the other models, but vectors corresponding to each point on the surface. It was trained for two cycles of 1,000 epochs each via transfer learning. The convolution-based models and PI-ConvTF were configured such that the past $10$ days were used to predict the data for the $11^{th}$ day. The configuration details of all models are listed in Table \ref{table:1}.

\begin{table}[htp]
    \centering
    \begin{tabular}{|c c c c c c c|}
         \hline
         Models & Epochs & Batch Size & Initial LR & LSTM Kernel Size & Hidden Channels & Attention Heads \\ 
         \hline\hline
         PINN & 2000 & 256 & 0.1 & - & - & - \\
         \hline
         ConvLSTM & 100 & 32 & 0.001 & 3x3 & 64 & - \\
         \hline
         SA-ConvLSTM & 100 & 32 & 0.001 & 3x3 & 64 & - \\
         \hline
         ConvTF & 100 & 16 & 0.001 & - & 32 & 4 \\
         \hline
         PI-ConvTF & 100 & 16 & 0.001 & - & 32 & 4 \\
         \hline
    \end{tabular}
    \captionsetup{skip=10pt}
    \caption{The configuration and hyperparameter details of the convolution-based models and PI-ConvTF.}
    \label{table:1}
\end{table}

Note that the architecture settings used here are those that recorded the peak performance according to our experiments. We conducted an empirical grid search over each hyperparameter, testing epochs in the range [50, 100, 150], batch sizes in [16, 32, 64], and initial learning rates in [1e-4, 0.001, 0.01]. For the PINN model, we used epochs in the range [500, 1000] for each training cycle and tested batch sizes of [128, 256]. For other hyperparameters that are more model-specific, we adhered to the configurations presented in the original papers where these architectures were introduced in \citep{shi2015convolutional}, \citep{lin2020selfattention}, \citep{liu2021convtransformer}. For the SA-ConvLSTM model, the query channels and key tensors were all set to eight, and the lambda value was 0.1 for PI-ConvTF. All models performed the best when using only a single layer. A learning rate scheduling was adopted based on the validation loss (also calculated by MAPE) decrease between epochs. Although the training was thoroughly conducted for the number of epochs denoted in Table \ref{table:1}, the weights for the result analysis were taken from the epoch with the lowest validation loss.

\subsection{Results and Analysis} \label{experiments}

\paragraph{Volatility Surface Prediction and Call Option Price Inference.}

\begin{figure}[htp]
    \centering
    \includegraphics[scale=0.85, width=\textwidth]{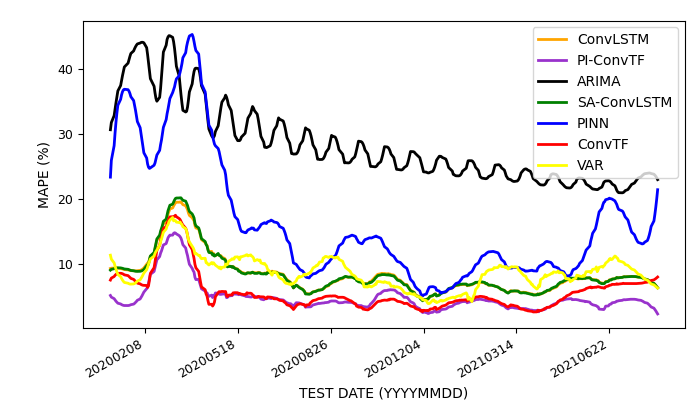}
    \caption{MAPEs for volatility surface prediction from five deep learning models, ARIMA, and VAR on test dates. The figure presents the outcomes from models trained without the exclusion of outliers from the training dataset.}
    \label{fig:TESTMAPE}
\end{figure}

Figure \ref{fig:TESTMAPE} shows the performance of the five models configured in Table \ref{table:1} {and VAR} {and ARIMA} on the daily volatility surface prediction task targeted on the test set and evaluated by the MAPE metric per day.

The advantage of deploying convolution-based architectures is clear. They can better estimate the statistical relationships between volatility surface points and between days than a physics-informed DNN structure. This is because of the nature of these models, which are based on convolution operations and the LSTM/transformer design, which are known for their efficacy in capturing spatial and temporal dependencies, respectively. Moreover, the effects of the attention mechanism and transformer architecture are visible from the decrease in the error shown by ConvTF.

As for the proposed PI-ConvTF architecture, performance improvement during the dates of historically high volatility is recognizable, which is similar to our test case during the initial days of the COVID-19 pandemic. This is because of the role of the physics-informed loss, which enables convolution-based models to learn relationships that obey the volatility surface's physics and additional statistical properties.

We compare the performance of our proposed method with standard statistical methods such as VAR and ARIMA. We have observed that ARIMA struggles to capture the dynamics of the surface. This is mainly due to its pointwise univariate prediction scheme, which lacks the necessary expressiveness to accurately describe the complexity of the task.
VAR, which uses multivariate autoregression, has stronger expressive power for modeling time series than the plain PINN model, which does not employ statistical tools to capture dependencies of volatility values over time.
However, the convolution-based models show comparable or improved prediction accuracy because their learning process is closely related to the spatial and temporal statistics of the volatility surface.
Overall, the transformer-based models tend to outperform VAR.

\begin{figure}[htp]
    \centering
    \includegraphics[scale=0.2, width=0.9\textwidth]{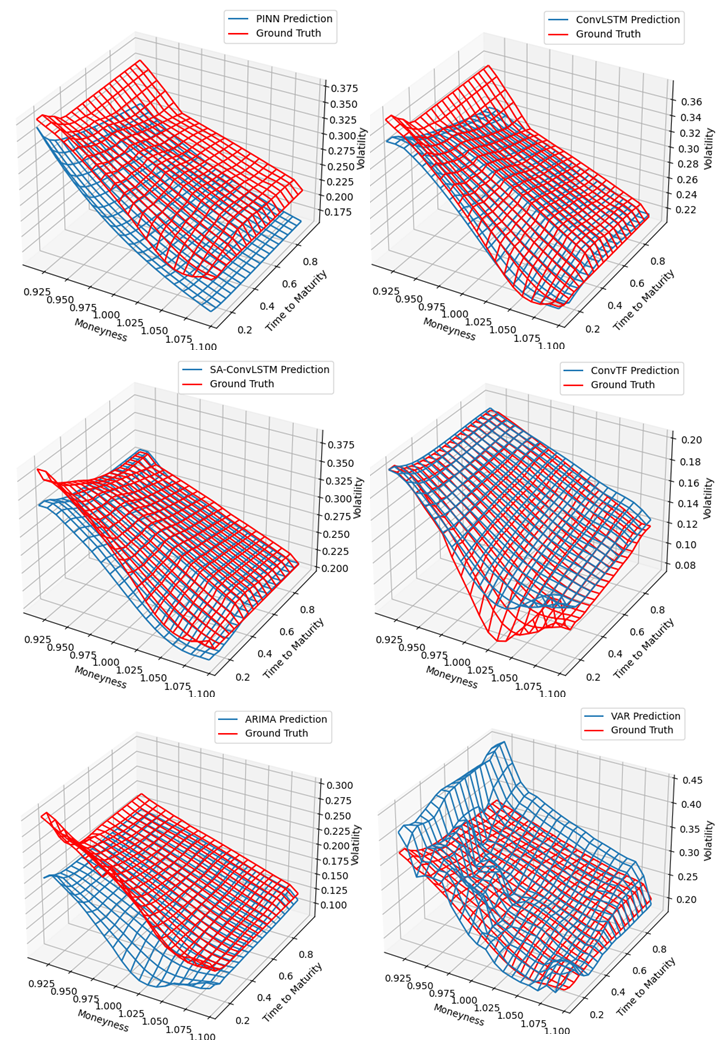}
    \caption{Predicted volatility surfaces using PINN, ConvLSTM, SA-ConvLSTM, ConvTF, ARIMA, and VAR. This figure presents a snapshot of the ground-truth and predicted volatility surfaces of each model for randomly sampled dates in the test data.}
    \label{fig:pred_vol_surf}
\end{figure}

\begin{figure}[htp]
    \centering
    \includegraphics[scale=0.35, width=\textwidth]{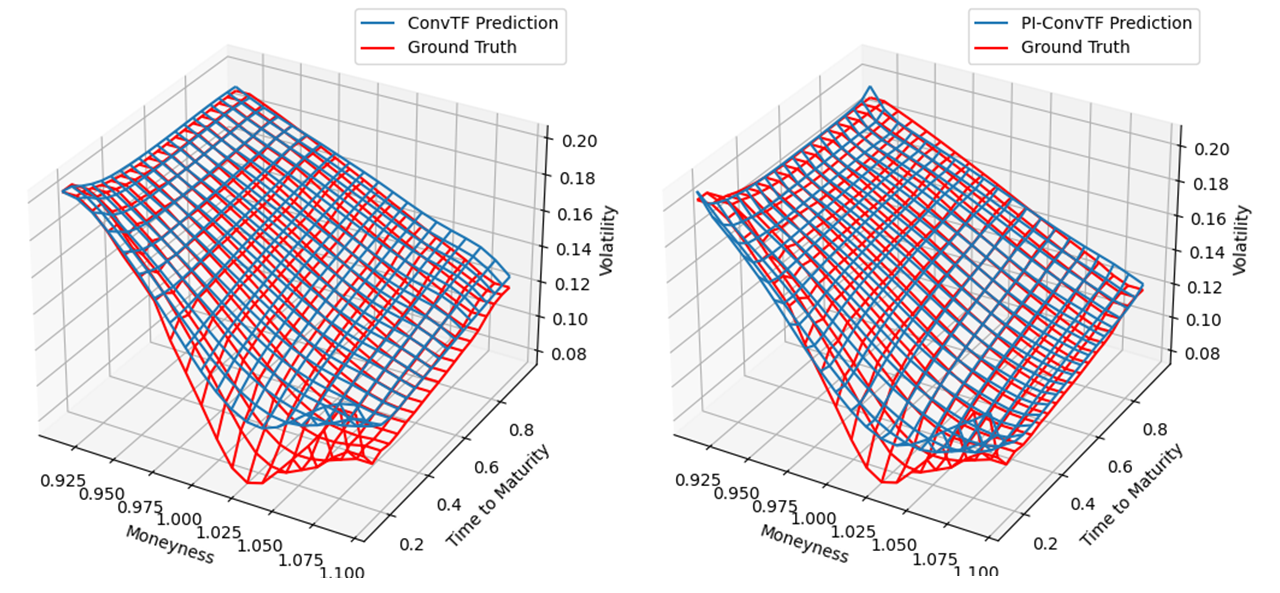}
    \caption{Daily snapshots of predicted volatility surfaces implemented by ConvTF and PI-ConvTF.}
    \label{fig:piconvtf_convtf_pred_vol_surf}
\end{figure}

{
Figure \ref{fig:pred_vol_surf} shows the predicted volatility surfaces for selected test dates using six different models. The PINN, VAR, and ARIMA models generally fail to capture changes in the surface and their predictions are either too high or too low.
While the ConvLSTM and SA-ConvLSTM models improve upon these shortcomings, they have problematic predictions at the boundary of the grid.
The ConvTF model provides robust performance, but sometimes gives incorrect predictions when volatility changes rapidly with respect to moneyness value, such as near the at-the-money (ATM) option.
The PI-ConvTF model, which incorporates the Black-Scholes equation as a physics-informed loss, is able to accurately predict rapid changes in the surface and mitigate this issue; see e.g. Figure \ref{fig:piconvtf_convtf_pred_vol_surf}.
}

\begin{figure}[htp]
    \centering
    \includegraphics[scale=0.35, width=0.9\textwidth]{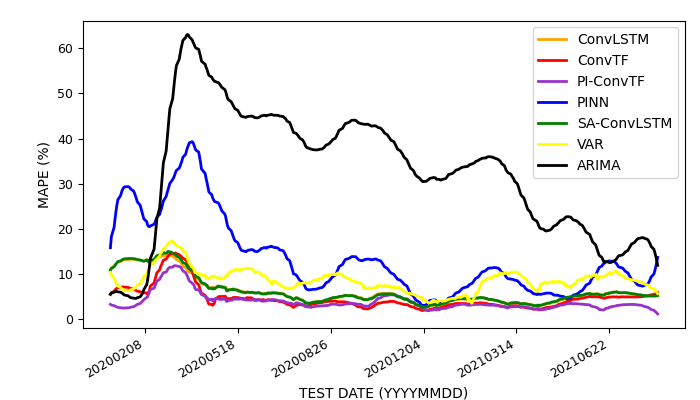}
    \caption{Daily MAPEs for the inferred call option price on test data for each model. The MAPE values were obtained after excluding the outliers below the $20^{th}$ percentile among the inferred prices per model.}
    \label{fig:call_TESTMAPE}
\end{figure}

To further understand the accuracy and practicality of the predicted volatility surfaces, we also infer the call option price for every point on the predicted surface for each model using Equations \eqref{e:piconvtf1} and \eqref{e:piconvtf2}. Owing to the characteristics of Equation \eqref{e:piconvtf1}, minor errors in the volatility value can translate into significant errors in the call option price evaluation. Therefore, to reasonably exclude outliers within the inferred call option prices, we excluded evaluated call option prices below the $20^{th}$ percentile for each day during MAPE calculation. Figure \ref{fig:call_TESTMAPE} shows the MAPE values calculated in this manner for each test day.

\begin{figure}[htp]
    \centering
    \includegraphics[scale=0.35, width=\textwidth]{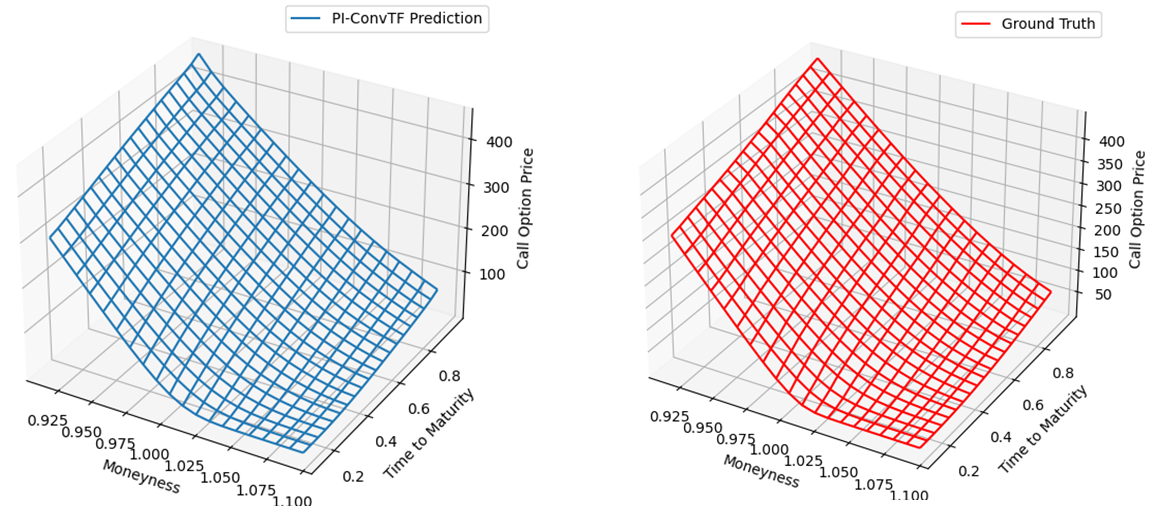}
    \caption{Comparison of the call option prices for a sample test day predicted by PI-ConvTF and the baseline call option prices.}
    \label{fig:call_pred_comp}
\end{figure}

\begin{figure}[htp]
    \centering
    \includegraphics[scale=0.35, width=\textwidth]{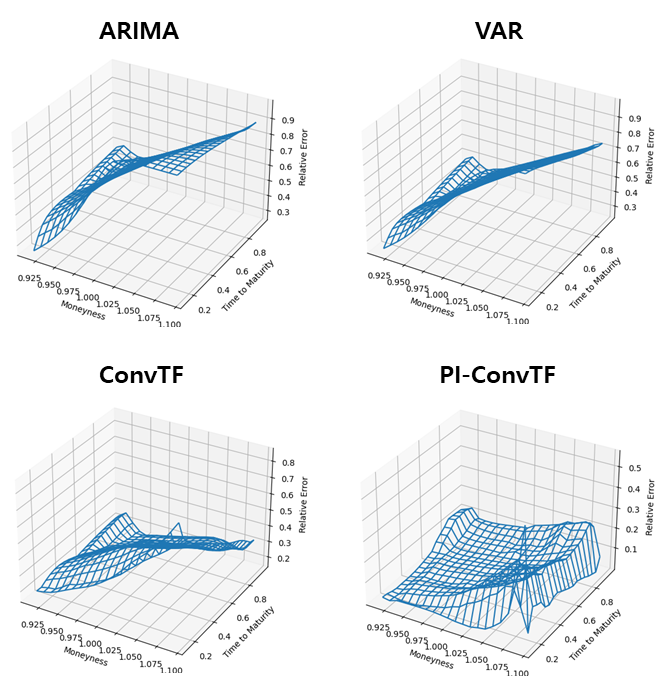}
    \caption{Relative errors of call option prices inferred from {ARIMA, VAR, ConvTF, and PI-ConvTF}. Note that this figure shows the errors from a sampled test day and without excluding outliers.}
    \label{fig:call_diffs}
\end{figure}

Although the changes in errors across the test dates in the call option prices are similar to those of our predicted volatility values most of the time, during days of high volatility (COVID-19 pandemic), the superior performance of PI-ConvTF is distinguishable from the convolution-based models. While ConvTF outperforms ConvLSTM and SA-ConvLSTM with more advanced statistical tools, such as attention to volatility surface prediction, their accuracies become more similar when it comes to call option price evaluation. This is because ConvTF does not consider the physics between volatility and call option price during training. In contrast, PI-ConvTF exploits this relationship, thereby inferring the call option price more accurately. {In Figure \ref{fig:call_diffs}, we present the relative errors between the baseline and predicted call option prices calculated using the Black-Scholes equation. The baseline call option prices are calculated using interpolated volatility values, while the predicted call option prices are derived from predicted volatility values obtained through ARIMA, VAR, ConvTF, and PI-ConvTF predictions.
The relative error is simply defined by $\|Call_{base} - Call_{pred}\| / Call_{base}$. It is worth noting that ARIMA and VAR tend to struggle in accurately predicting the dynamics of call option prices with respect to moneyness and time to maturity. Furthermore, the comparison of volatility surface predictions in Figure  \ref{fig:piconvtf_convtf_pred_vol_surf} demonstrates that PI-ConvTF outperforms ConvTF in predicting the prices of ATM call options.} 
A comparison between the call option prices predicted by the PI-ConvTF and baseline prices is illustrated in Figure \ref{fig:call_pred_comp}.
We easily see that the option prices are very close to each other.

Table \ref{table:2} presents the MAPE values of the five models {and VAR} {and ARIMA }averaged across the test dates. It is worth noting that the high average error from ARIMA is partly due to the relatively frequent outliers present in volatile test dates. Otherwise, the model performances on volatility prediction are as expected. We can also see that SA-ConvLSTM slightly underperforms ConvLSTM in both volatility and call option price inference in terms of average error values. 

\begin{table}[htp]
    \centering
    \begin{tabular}{|c c c|}
         \hline
         Models & Volatility MAPE (\%) & Call Price MAPE (\%) \\ 
         \hline\hline
         PINN & 16.4823 & 13.5132 \\
         \hline
         ConvLSTM & 8.4060 & 6.0774 \\
         \hline
         SA-ConvLSTM & 8.4218 & 6.1265 \\
         \hline
         ConvTF & 5.7891 & 4.5944 \\
         \hline
         PI-ConvTF & 4.9174 & 3.8457 \\
         \hline
         VAR & 8.8043 & 8.5821 \\
         \hline
         ARIMA & 28.2453 & 32.5921 \\
         \hline
    \end{tabular}
    \captionsetup{skip=10pt}
    \caption{Average MAPE values across test dates for all models trained with outlier data. Call price MAPE values are determined after excluding outlier values below the $20^{th}$ percentile.}
    \label{table:2}
\end{table}

\begin{figure}[htp]
    \centering
    \includegraphics[scale=0.35, width=\textwidth]{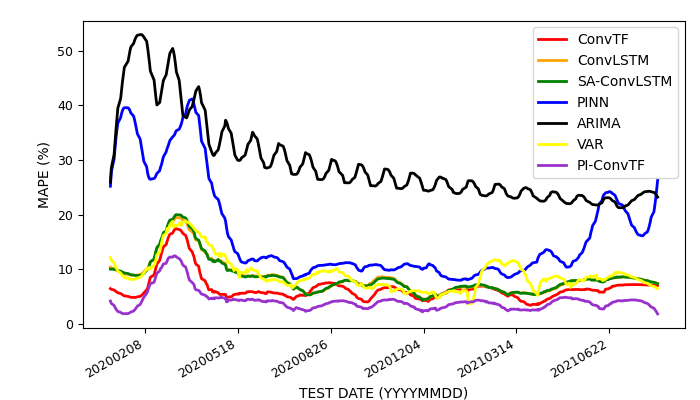}
    \caption{Daily MAPEs for volatility prediction on test data for each model trained without outlier data.}
    \label{fig:no-outliers_vol_TESTMAPE}
\end{figure}

\begin{figure}[htp]
    \centering
    \includegraphics[scale=0.35, width=\textwidth]{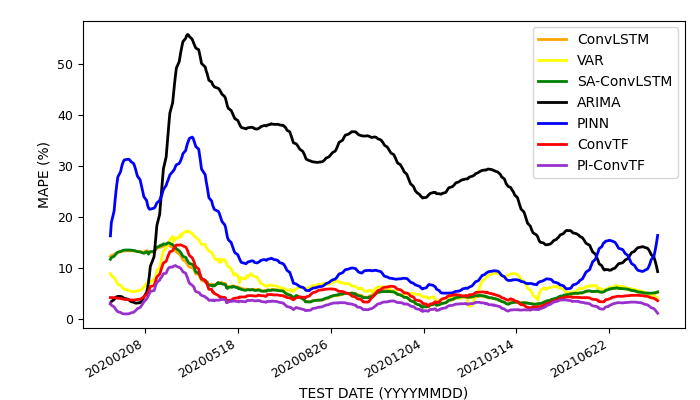}
    \caption{Daily MAPEs for the inferred call option price on test data for each model trained without outlier data. The MAPE values were obtained after excluding the outliers below the $20^{th}$ percentile among the inferred prices per model.}
    \label{fig:no-outliers_call_TESTMAPE}
\end{figure}

\paragraph{Exclusion of Outliers in the Training Data.}
In our results, outliers in inferred call option prices frequently appear when estimating call options with small values, particularly around short maturities and large strikes. This anomaly is attributed to the errors in volatility prediction amplifying for call prices below \$1. In simpler terms, even minor discrepancies in volatility prediction lead to significant errors when the call option price is deduced via the Black-Scholes equation. To further examine the impact of these outliers, we retrained our models, excluding data points with call option prices below the  $20^{th}$ percentile for each specific day in the training set. We maintained all other configurations unchanged. The outcomes of this adjustment are documented in Table \ref{table:no-outliers} and Figures \ref{fig:no-outliers_vol_TESTMAPE}, \ref{fig:no-outliers_call_TESTMAPE}. A review of these results suggests that omitting outliers during training yields slight improvements in accuracy for most models compared to models trained with outlier data.

\begin{table}[htp]
    \centering
    \begin{tabular}{|c c c|}
         \hline
         Models & Volatility MAPE (\%) & Call Price MAPE (\%) \\ 
         \hline\hline
         PINN & 16.4585 & 12.7439 \\
         \hline
         ConvLSTM & 8.4399 & 6.0522 \\
         \hline
         SA-ConvLSTM & 8.4491 & 6.0814 \\
         \hline
         ConvTF & 6.5473 & 5.0863 \\
         \hline
         PI-ConvTF & 4.1982 & 3.2844 \\
         \hline
         VAR & 9.1451 & 7.2821 \\
         \hline
         ARIMA & 29.4165 & 26.8199 \\
         \hline
    \end{tabular}
    \captionsetup{skip=10pt}
    \caption{MAPE values averaged across test dates of all models trained without outlier data. Call price MAPE values are calculated after removing outlier values below the $20^{th}$ percentile.}
    \label{table:no-outliers}
\end{table}

In order to leverage the put-call parity relationship and study the put option prices inferred from our volatility prediction models, we employ the monthly dividends data of the S\&P 500. This allows us to compute put prices based on calls with corresponding maturity and strike pairs.
Drawing from \citep{hull2003textbook}, we reference the following equation:

\begin{equation} 
\begin{split}
    C + Ke^{-r\tau} = P + S - \sum_{i=1}^{n} D_{i}e^{-rt_{i}}
\end{split} \label{e:put-call-parity}
\end{equation}

where $C$ and $P$ are call and put option prices and $K$, $r$, $\tau$, $S$ are associated strike, risk-free rate, time to maturity, and underlying stock price, respectively. $D_{i}$, $t_{i}$ represent the dividend paid at, and the time to reach the $i^{th}$ month. 
We estimate the call price using our models trained without outliers, and then deduce the put price based on equation \ref{e:put-call-parity}. The experimental results are presented in Table \ref{table:put_MAPE}. Notably, these results are consistent with the call option inference in terms of the models' prediction capability.

\begin{table}[htp]
    \centering
    \begin{tabular}{|c c|}
         \hline
         Models & Put Price MAPE (\%) \\ 
         \hline\hline
         PINN & 10.8237 \\
         \hline
         ConvLSTM & 4.0159 \\
         \hline
         SA-ConvLSTM & 4.0352 \\
         \hline
         ConvTF & 4.8325 \\
         \hline
         PI-ConvTF & 2.9297 \\
         \hline
         VAR & 6.7802 \\
         \hline
         ARIMA & 23.4107 \\
         \hline
    \end{tabular}
    \captionsetup{skip=10pt}
    \caption{Average put option price MAPE values across test dates for all models trained without outlier data. These inferred put option prices are derived from the call option prices previously estimated by our models. The MAPE values are computed after excluding outlier values of put options that fall below the  $20^{th}$ percentile.}
    \label{table:put_MAPE}
\end{table}

\paragraph{Exclusion of Volatile Regimes during Training.}

\begin{figure}[htp]
    \centering
    \includegraphics[scale=0.8]{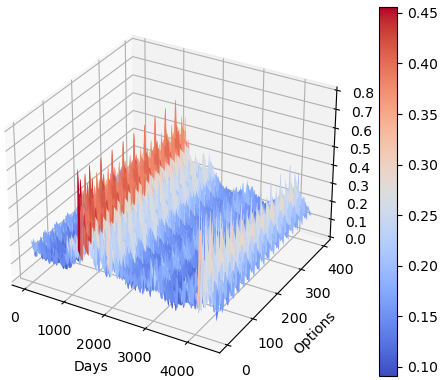}
    \caption{Distribution of implied volatility values from their associated call options across time.}
    \label{fig:vol_dist}
\end{figure}

To test whether our models perform well in making predictions during historically volatile regimes when they have not been trained on similar data, we first analyze the distribution of volatility data. Figure \ref{fig:vol_dist} shows the volatility values across the total number of dates and per 400 options (strike and maturity pairs) for each date. 

\begin{table}[htp]
    \centering
    \begin{tabular}{|c c c|}
         \hline
         Target Test Dates & Training Period & Test Period \\ 
         \hline\hline
         Subprime Mortgage Crisis & 2004/1/5 - 2008/9/25 & 2008/9/26 - 2009/5/11 \\
         \hline
         Initial Days of the COVID-19 Pandemic & 2009/5/12 - 2020/3/4 & 2020/3/5 - 2020/4/21 \\
         \hline
    \end{tabular}
    \captionsetup{skip=10pt}
    \caption{Additional train and test data splits for further experiments with two noticeable periods as volatile regimes: the breakout of the Subprime Mortgage Crisis and the initial days of the COVID-19 pandemic.}
    \label{table:3}
\end{table}

As shown in Figure \ref{fig:vol_dist}, two noticeable periods can be classified as volatile regimes: the breakout of the Subprime Mortgage Crisis and the initial days of the COVID-19 pandemic. Calculations show that the $95^{th}$ percentile of this distribution is approximately $0.3093$. The dates when the mean of all option volatilities exceeds this statistic are between 2008/9/26 and 2009/5/11 and between 2020/3/5 and 2020/4/21. Based on this finding, we retrained our models on two additional training, validation, and test splits of data, as shown in Table \ref{table:3}. The validation periods were set similar to the latest 20\% of the training period.

\begin{table}[htp]
    \centering
    \begin{tabular}{|c c c|}
         \hline
         Models & Subprime Crisis MAPE (\%) & COVID-19 MAPE (\%) \\ 
         \hline\hline
         PINN & 30.5441 & 31.3940 \\
         \hline
         ConvLSTM & 20.3654 & 16.9395 \\
         \hline
         SA-ConvLSTM & 22.6122 & 17.3269 \\
         \hline
         ConvTF & 66.5141 & 21.3854 \\
         \hline
         PI-ConvTF & 65.2277 & 22.6640 \\
         \hline 
    \end{tabular}
    \captionsetup{skip=10pt}
    \caption{MAPE averaged across test dates for new data splits using the breakout of the Subprime Mortgage Crisis and the initial days of the COVID-19 pandemic.}
    \label{table:4}
\end{table}

\begin{figure}[htp]
    \centering
    \includegraphics[scale=0.35, width=\textwidth]{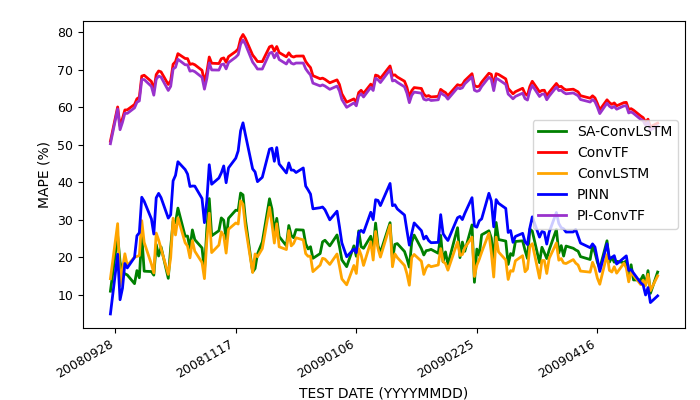}
    \caption{Daily MAPE of volatility predictions on the Subprime Crisis test period.}
    \label{fig:subprime_TESTMAPE}
\end{figure}

\begin{figure}[htp]
    \centering
    \includegraphics[scale=0.35, width=\textwidth]{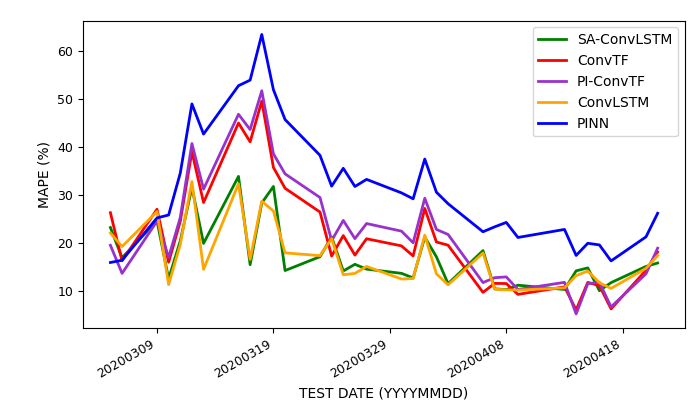}
    \caption{Daily MAPE of volatility predictions on the COVID-19 test period.}
    \label{fig:covid_TESTMAPE}
\end{figure}

Table \ref{table:4} presents the MAPE values evaluated during the test period (averaged across the test dates) for the two training cycles conducted on the new data splits. An in-depth view of the MAPE of the daily predictions for the two new test periods per model is provided in Figures \ref{fig:subprime_TESTMAPE} and \ref{fig:covid_TESTMAPE}.

The main difference between training and testing under our new conditions is the notable decrease in performance of the transformer-based models, that is, ConvTF and PI-ConvTF. This seems to be because transformer architectures are inherently more complex than LSTM architectures, thereby having a higher risk of overfitting if not given sufficiently variant data. As our experiments intentionally excluded the training of volatile regimes but conducted tests on highly volatile data, the simpler architectures of ConvLSTM and SA-ConvLSTM outperformed ConvTF and PI-ConvTF. However, as shown in Table \ref{table:4}, an overall performance decrease was commonly observed in all models, and testing on the days of the Subprime Mortgage Crisis showed worse results as the training period itself was much shorter. 

It would be interesting to further experiment with variants of PI-ConvTF under these particular training and test splits, such as adopting ConvLSTM and SA-ConvLSTM, which showed optimal performance, as the baseline architecture with the Black-Scholes equation. In addition, as the lookback timestep window was fixed at 10 in our case, observing whether transformer-based architectures perform better than LSTM-based models with increased timestep sizes can also be considered a future task.

\section{Conclusion}
In this study, we compare the physical and statistical deep-learning approaches to the prediction of the volatility surface of options. The standard PINN with a DNN architecture was utilized for the physical approach, whereas three convolution-based architectures: ConvLSTM, SA-ConvLSTM, and ConvTF, were used for the statistical approach. The results show that statistical predictions generally perform better in this specific task and layered structures yield more significant errors. We further propose the PI-ConvTF architecture by replacing the DNN in the standard PINN with ConvTF and performing the required operations to adopt the Black-Scholes equation in the loss function. PI-ConvTF successfully benefits from its physical and statistical components, as the two methods reinforce each other to achieve higher accuracies than in the case of individual predictions. Further experiments to evaluate the call option price from predicted volatilities show the efficacy of PI-ConvTF, while studying the changes in model performance under different training and test data settings proves that PI-ConvTF suffers from the problem of overfitting because of architecture complexity when not given sufficiently variant data for training.

\section*{Declaration of Interest Statement}
The authors declare that they have no known competing financial interests or personal relationships that could have appeared to influence the work reported in this paper.

\bigskip

\section*{Data Availability Statement}
The authors confirm that the data supporting the findings of this study are available within the article.

\section*{Acknowledgment}
The authors thank the anonymous referees for their helpful comments that improved the quality of the manuscript.
\typeout{}
\bibliographystyle{unsrtnat}
\bibliography{references}

\begin{thebibliography}{42}
\providecommand{\natexlab}[1]{#1}
\providecommand{\url}[1]{\texttt{#1}}
\expandafter\ifx\csname urlstyle\endcsname\relax
  \providecommand{\doi}[1]{doi: #1}\else
  \providecommand{\doi}{doi: \begingroup \urlstyle{rm}\Url}\fi

\bibitem[Black and Scholes(1972)]{black1972the}
Fischer~S. Black and Myron~S. Scholes.
\newblock The valuation of option contracts and a test of market efficiency.
\newblock \emph{Journal of Finance}, 27\penalty0 (2):\penalty0 399--417, 1972.

\bibitem[Black and Scholes(1973)]{black1973the}
Fischer~S. Black and Myron~S. Scholes.
\newblock The pricing of options and corporate liabilities.
\newblock \emph{The Journal of Political Economy}, 81\penalty0 (3):\penalty0
  637--654, 1973.

\bibitem[Merton(1973)]{merton1973theory}
Robert~C. Merton.
\newblock Theory of rational option pricing.
\newblock \emph{The Bell Journal of Economics and Management Science},
  4\penalty0 (1):\penalty0 141--183, 1973.

\bibitem[Garman and Kohlhagen(1983)]{garman1983foreign}
Mark~B. Garman and Steven~W. Kohlhagen.
\newblock Foreign currency option values.
\newblock \emph{Journal of International Money and Finance}, 2\penalty0
  (3):\penalty0 231--237, 1983.

\bibitem[Shinde and Takale(2012)]{shinde2012study}
A.~S. Shinde and K.~C. Takale.
\newblock Study of black-scholes model and its applications.
\newblock \emph{Procedia Engineering}, 38:\penalty0 270--279, 2012.

\bibitem[Hull and White(1987)]{hull1987thepricing}
J.~C. Hull and A.~White.
\newblock The pricing of options on assets with stochastic volatilities.
\newblock \emph{Journal of Finance}, 42:\penalty0 281--300, 1987.

\bibitem[Heston(1993)]{heston1993aclosed}
Steven~L. Heston.
\newblock A closed-form solution for options with stochastic volatility with
  applications to bond and currency options.
\newblock \emph{Review of Financial Studies}, 6\penalty0 (2):\penalty0
  327--343, 1993.

\bibitem[Derman and Kani(1994)]{derman1994riding}
Emanuel Derman and Iraj Kani.
\newblock Riding on a smile.
\newblock \emph{Risk}, 7\penalty0 (2):\penalty0 32--39, 1994.

\bibitem[Cont and Da~Fonseca(2002)]{CD02}
Rama Cont and Jose Da~Fonseca.
\newblock Dynamics of implied volatility surfaces.
\newblock \emph{Quantitative Finance}, 2\penalty0 (1):\penalty0 45--60, 2002.

\bibitem[Jiang and Tian(2005)]{JT05}
George~J. Jiang and Yisong~S. Tian.
\newblock The model-free implied volatility and its information content.
\newblock \emph{The Review of Financial Studies}, 18\penalty0 (4):\penalty0
  1305--1342, 2005.
\newblock ISSN 08939454, 14657368.
\newblock URL \url{http://www.jstor.org/stable/3598022}.

\bibitem[Dupire(1994)]{dupire1994pricing}
Bruno Dupire.
\newblock Pricing with a smile.
\newblock \emph{Risk}, 7\penalty0 (1):\penalty0 18--20, 1994.

\bibitem[Granger and Poon(2003)]{granger2003forecasting}
Clive W.~J. Granger and Ser-Huang Poon.
\newblock Forecasting volatility in financial markets: A review.
\newblock \emph{Journal of Economic Literature}, 41\penalty0 (2):\penalty0
  478--539, 2003.

\bibitem[Avellaneda et~al.(1997)Avellaneda, Friedman, Holmes, and
  Samperi]{avellaneda1997calib}
M.~Avellaneda, C.~Friedman, R.~Holmes, and D.~Samperi.
\newblock Calibrating volatility surfaces via relative-entropy minimization.
\newblock \emph{Applied Mathematical Finance}, 4\penalty0 (1):\penalty0 37--64,
  1997.

\bibitem[Andersen and Brotherton-Ratcliffe(1998)]{andersen1998the}
L.~Andersen and R.~Brotherton-Ratcliffe.
\newblock The equity option volatility smile: an implicit finite difference
  approach.
\newblock \emph{The Journal of Computational Finance}, 1:\penalty0 5--32, 1998.

\bibitem[Berestycki et~al.(2002)Berestycki, Busca, and Florent]{calib}
H.~Berestycki, J.~Busca, and I.~Florent.
\newblock Asymptotics and calibration of local volatility models.
\newblock \emph{Quantitative Finance}, 2\penalty0 (61), 2002.

\bibitem[Bondarenko and Bondarenko(2018)]{bondarenko2018calib}
Maksym Bondarenko and Victor Bondarenko.
\newblock Calibration of dupire local volatility model using genetic algorithm
  of optimization.
\newblock \emph{Neuro-Fuzzy Modeling Techniques in Economics}, 7\penalty0
  (1):\penalty0 1--20, 2018.

\bibitem[Mixon(2002)]{factors}
S.~Mixon.
\newblock Factors explaining movements in the implied volatility surface.
\newblock \emph{The Journal of Futures Markets}, 22\penalty0 (10), 2002.

\bibitem[Noh et~al.(1994)Noh, Engle, and Kane]{linreg}
J.~Noh, R.F. Engle, and A.~Kane.
\newblock Forecasting volatility and option prices of the s\&p 500 index.
\newblock \emph{The Journal of Derivatives}, 2, 1994.

\bibitem[Malliaris and Salchenberger(1996)]{malliaris1996using}
Mary Malliaris and Linda Salchenberger.
\newblock Using neural networks to forecast the s\&p 100 implied volatility.
\newblock \emph{Neurocomputing}, 10:\penalty0 183--195, 1996.

\bibitem[Raissi et~al.(2019)Raissi, Perdikaris, and
  Karniadakis]{raissi2019physics}
M.~Raissi, P.~Perdikaris, and G.E. Karniadakis.
\newblock Physics-informed neural networks: A deep learning framework for
  solving forward and inverse problems involving nonlinear partial differential
  equations.
\newblock \emph{Journal of Computational Physics}, 378:\penalty0 686--707,
  2019.

\bibitem[Ameya et~al.(2020)Ameya, Jagtap, , 9597, , Jagtap, George,
  Karniadakis, , 9598, , and Karniadakis]{PINN001}
Ameya, D.~Jagtap, , 9597, , Ameya~D. Jagtap, George, Em~Karniadakis, , 9598, ,
  and George~Em Karniadakis.
\newblock Extended physics-informed neural networks (xpinns): A generalized
  space-time domain decomposition based deep learning framework for nonlinear
  partial differential equations.
\newblock \emph{Communications in Computational Physics}, 28\penalty0
  (5):\penalty0 2002--2041, 2020.
\newblock ISSN 1991-7120.

\bibitem[Raissi et~al.(2020)Raissi, Yazdani, and Karniadakis]{PINN002}
Maziar Raissi, Alireza Yazdani, and George~Em Karniadakis.
\newblock Hidden fluid mechanics: Learning velocity and pressure fields from
  flow visualizations.
\newblock \emph{Science}, 367\penalty0 (6481):\penalty0 1026--1030, 2020.

\bibitem[Karniadakis et~al.(2021)Karniadakis, Kevrekidis, Lu, Perdikaris, Wang,
  and Yang]{PINN003}
George~Em Karniadakis, Ioannis~G. Kevrekidis, Lu~Lu, Paris Perdikaris, Sifan
  Wang, and Liu Yang.
\newblock Physics-informed machine learning.
\newblock \emph{Nature Reviews Physics}, 3\penalty0 (6):\penalty0 422--440,
  2021.

\bibitem[Mathews et~al.(2021)Mathews, Francisquez, Hughes, Hatch, Zhu, and
  Rogers]{PINNa-001}
A.~Mathews, M.~Francisquez, J.~W. Hughes, D.~R. Hatch, B.~Zhu, and B.~N.
  Rogers.
\newblock Uncovering turbulent plasma dynamics via deep learning from partial
  observations.
\newblock \emph{Phys. Rev. E}, 104:\penalty0 025205, Aug 2021.
\newblock \doi{10.1103/PhysRevE.104.025205}.
\newblock URL \url{https://link.aps.org/doi/10.1103/PhysRevE.104.025205}.

\bibitem[Wiecha et~al.(2021)Wiecha, Arbouet, Girard, and Muskens]{PINNa-002}
Peter~R. Wiecha, Arnaud Arbouet, Christian Girard, and Otto~L. Muskens.
\newblock Deep learning in nano-photonics: inverse design and beyond.
\newblock \emph{Photon. Res.}, 9\penalty0 (5):\penalty0 B182--B200, May 2021.

\bibitem[Li et~al.(2021{\natexlab{a}})Li, Bazant, and Zhu]{PINNa-003}
Wei Li, Martin~Z. Bazant, and Juner Zhu.
\newblock A physics-guided neural network framework for elastic plates:
  Comparison of governing equations-based and energy-based approaches.
\newblock \emph{Computer Methods in Applied Mechanics and Engineering},
  383:\penalty0 113933, 2021{\natexlab{a}}.
\newblock ISSN 0045-7825.

\bibitem[Kissas et~al.(2020)Kissas, Yang, Hwuang, Witschey, Detre, and
  Perdikaris]{PINNa-004}
Georgios Kissas, Yibo Yang, Eileen Hwuang, Walter~R. Witschey, John~A. Detre,
  and Paris Perdikaris.
\newblock Machine learning in cardiovascular flows modeling: Predicting
  arterial blood pressure from non-invasive 4d flow mri data using
  physics-informed neural networks.
\newblock \emph{Computer Methods in Applied Mechanics and Engineering},
  358:\penalty0 112623, 2020.
\newblock ISSN 0045-7825.

\bibitem[Shi et~al.(2015)Shi, Chen, Wang, Yeung, Wong, and
  Woo]{shi2015convolutional}
Xingjian Shi, Zhourong Chen, Hao Wang, Dit-Yan Yeung, Wai-kin Wong, and
  Wang-chun Woo.
\newblock Convolutional lstm network: A machine learning approach for
  precipitation nowcasting.
\newblock In \emph{NIPS 2015}, pages 802--810, 2015.

\bibitem[Azad et~al.(2019)Azad, Asadi-Aghbolaghi, Fathy, and
  Escalera]{convlstm001}
Reza Azad, Maryam Asadi-Aghbolaghi, Mahmood Fathy, and Sergio Escalera.
\newblock Bi-directional convlstm u-net with densley connected convolutions.
\newblock In \emph{Proceedings of the IEEE/CVF International Conference on
  Computer Vision (ICCV) Workshops}, Oct 2019.

\bibitem[Moishin et~al.(2021)Moishin, Deo, Prasad, Raj, and
  Abdulla]{convlstm002}
Mohammed Moishin, Ravinesh~C. Deo, Ramendra Prasad, Nawin Raj, and Shahab
  Abdulla.
\newblock Designing deep-based learning flood forecast model with convlstm
  hybrid algorithm.
\newblock \emph{IEEE Access}, 9:\penalty0 50982--50993, 2021.
\newblock \doi{10.1109/ACCESS.2021.3065939}.

\bibitem[Lin et~al.(2020)Lin, Li, Zheng, Cheng, and Yuan]{lin2020selfattention}
Zhihui Lin, Maomao Li, Zhuobin Zheng, Yangyang Cheng, and Chun Yuan.
\newblock Self-attention convlstm for spatiotemporal prediction.
\newblock \emph{Proceedings of the AAAI Conference on Artificial Intelligence},
  34\penalty0 (7):\penalty0 11531--11538, 2020.

\bibitem[Li et~al.(2021{\natexlab{b}})Li, Tang, Deng, and Zhao]{sa-lstm-01}
Biao Li, Baoping Tang, Lei Deng, and Minghang Zhao.
\newblock Self-attention convlstm and its application in rul prediction of
  rolling bearings.
\newblock \emph{IEEE Transactions on Instrumentation and Measurement},
  70:\penalty0 1--11, 2021{\natexlab{b}}.
\newblock \doi{10.1109/TIM.2021.3086906}.

\bibitem[Liu et~al.(2021)Liu, Luo, Li, Lu, Wu, Sun, Li, and
  Yang]{liu2021convtransformer}
Zhouyong Liu, Shun Luo, Wubin Li, Jingben Lu, Yufan Wu, Shilei Sun, Chunguo Li,
  and Luxi Yang.
\newblock Convtransformer: A convolutional transformer network for video frame
  synthesis.
\newblock \emph{arXiv preprint arXiv:2011.10185v2}, 2021.

\bibitem[Huang et~al.(2020)Huang, Hu, Yeung, and Chen]{convtf01}
Wenyong Huang, Wenchao Hu, Yu~Ting Yeung, and Xiao Chen.
\newblock Conv-transformer transducer: Low latency, low frame rate, streamable
  end-to-end speech recognition.
\newblock \emph{Proc. Interspeech 2020}, pages 5001--5005, 2020.

\bibitem[Hull(2003)]{hull2003textbook}
J.~Hull.
\newblock \emph{Options, Futures, and Other Derivatives}.
\newblock Prentice Hall, 2003.

\bibitem[Anwar and Andallah(2018)]{anwar2018astudy}
Nurul Anwar and Laek~Sazzad Andallah.
\newblock A study on numerical solution of black-scholes model.
\newblock \emph{Journal of Mathematical Finance}, 8:\penalty0 372--381, 2018.

\bibitem[Guo et~al.(2018)Guo, Loeper, and Wang]{guo2018local}
Ivan Guo, Gregoire Loeper, and Shiyi Wang.
\newblock Local volatility calibration by optimal transport.
\newblock \emph{arXiv preprint arXiv:1709.08075v4}, 2018.

\bibitem[Jin et~al.(2018)Jin, Wang, Kim, Heo, Yoo, Kim, Kim, and
  Jeong]{jin2018reconstruction}
Yuzi Jin, Jian Wang, Sangkwon Kim, Youngjin Heo, Changwoo Yoo, Youngrock Kim,
  Junseok Kim, and Darae Jeong.
\newblock Reconstruction of the time-dependent volatility function using the
  black-scholes model.
\newblock \emph{Discrete Dynamics in Nature and Society}, 2018:\penalty0 1--9,
  2018.

\bibitem[Cen and Le(2011)]{cen2011arobust}
Zhongdi Cen and Anbo Le.
\newblock A robust and accurate finite difference method for a generalized
  black-scholes equation.
\newblock \emph{Journal of Computational and Applied Mathematics}, 235\penalty0
  (13):\penalty0 3728--3733, 2011.

\bibitem[Lagaris et~al.(1998)Lagaris, Likas, and
  Fotiadis]{lagaris1998artificial}
Isaac~E. Lagaris, Aristidis Likas, and Dimitrios~I. Fotiadis.
\newblock Artificial neural networks for solving ordinary and partial
  differential equations.
\newblock \emph{IEEE Transactions on Neural Networks}, 9\penalty0 (5):\penalty0
  987--1000, 1998.

\bibitem[Kanniainen et~al.(2014)Kanniainen, Lin, and Yang]{tablestats}
Juho Kanniainen, Binghuan Lin, and Hanxue Yang.
\newblock Estimating and using garch models with vix data for option valuation.
\newblock \emph{Journal of Banking \& Finance}, 43:\penalty0 200--211, 2014.

\bibitem[de~Boor(1978)]{Boor1978APG}
Carl de~Boor.
\newblock A practical guide to splines.
\newblock In \emph{Applied Mathematical Sciences}, 1978.

\end{thebibliography}

\section*{Appendix}
{
The Black-Scholes equation with a nonconstant volatility is as follows: 

\begin{equation} \label{e:bse}
\begin{split}
     rC - \frac{\partial{C}}{\partial{t}} - rS\frac{\partial{C}}{\partial{S}} - \frac{1}{2}\sigma^2S^2\frac{\partial{^2C}}{\partial{S^2}} = 0,
\end{split}
\end{equation}

where $C:= C(S, t)$ is the call option price, $S$ the underlying asset price, $r$ the risk-free interest rate, and $\sigma:= \sigma(S, t)$ the volatility function. When the volatility function is constant, this equation is reduced to the classical Black-Scholes equation with a constant volatility, where the analytical solution is known to be

\begin{equation} 
\begin{split}
    C:= C_{BS} := S\Phi(d_{1}) - Ke^{-r(T-t)}\Phi(d_{2}).
\end{split} \label{e:bssol}
\end{equation}
Here, implied volatility is obtained by solving $C_{BS} = C_M$, where $C_M$ is an option price observed in the market.  
Note that $\Phi$ is the cumulative distribution function of the normal distribution, $T$ is the maturity, and

\begin{equation} 
\begin{split}
    & d_{1} = \frac{ln(S/K) + (r + \frac{1}{2}\sigma^2)(T-t)}{\sigma\sqrt{T - t}},\\ 
    & d_{2} = d_{1} - \sigma\sqrt{(T-t)}.
\end{split} \label{e:bssol2}
\end{equation}
}

\end{document}